\documentclass[prd,aps,showpacs,preprintnumbers,amsmath, amssymb]{revtex4-2}
\usepackage{amsmath,amssymb,graphics,epsfig,subfigure}
\usepackage{color}
\usepackage{multirow}
\usepackage{graphicx}
\usepackage{epstopdf}
\usepackage{dcolumn}
\usepackage{amsmath, mathrsfs}
\usepackage{amsmath}
\usepackage{amssymb}
\usepackage{bm}
\usepackage{color}
\usepackage{multirow}
\usepackage{float}
\usepackage[dvipsnames]{xcolor}
\usepackage{hyperref}
\UseRawInputEncoding
\begin{document}

\newcommand {\nn} {\nonumber}
\renewcommand{\baselinestretch}{1.3}

 \baselineskip=0.8cm
\title{Images of Braneworld black holes with radiatively inefficient accretion flows}

\author{Chengjia Chen$^{1}$\footnote{ccj@hunnu.edu.cn}, Yin Hao$^{1}$, Zelin Zhang$^{2}$, Qiyuan Pan$^{1}$\footnote{panqiyuan@hunnu.edu.cn}, Jiliang Jing$^{1}$ \footnote{jljing@hunnu.edu.cn}}


\affiliation{$^1$Department of Physics, Institute of Interdisciplinary Studies, Key Laboratory of Low Dimensional Quantum Structures
    and Quantum Control of Ministry of Education, Synergetic Innovation Center for Quantum Effects and Applications, and Hunan Research Center of the Basic Discipline for Quantum Effects and Quantum Technologies, Hunan Normal University,  Changsha, Hunan 410081, People's Republic of China
    \\
    $^2$Institute of Fundamental Physics and Quantum Technology, School of Physical Science and Technology, Ningbo University, Ningbo, Zhejiang 315211, P. R. China}

\begin{abstract}
 \baselineskip=0.6cm
Horizon-scale imaging acts as a transformative tool for probing spacetime geometry, enabling stringent tests of gravitational theories in the strong-field regime. The Casadio-Fabbri-Mazzacurati (CFM) black hole in braneworld contains an extra parameter that characterizes the tidal effects from the bulk geometry, making it highly valuable for this task. We perform  general relativistic radiative transfer (GRRT) simulations and generate synthetic images consistent with Event Horizon Telescope observations of M87*. We find that the tidal parameter imprints non-monotonic changes on the image morphology, underscoring the intricate coupling between spacetime geometry and the observable radiation from the accreting plasma. 
We also analyze the image-comparison metric using normalized cross-correlation coefficients and the DSSIM index and find that the magnitudes of these mismatches are on the order of $10^{-3}$, which implies that identifying braneworld black holes through black hole images remains challenging even with future ngEHT and BHEX observations.
\end{abstract}

\maketitle

\newpage
\section{Introduction}
\label{secIntroduction}

Einstein's general relativity (GR) is widely recognized as the most successful theory of gravity, having passed numerous observational and experimental tests with high precision \cite{EventHorizonTelescope:2019uob,EventHorizonTelescope:2022wkp,EventHorizonTelescope:2019ggy,EventHorizonTelescope:2022urf,LIGOScientific:2016aoc,Do:2019txf,Will:2014kxa}. Despite its remarkable achievements, GR is known to encounter fundamental theoretical and observational challenges, such as the spacetime singularity problem and the nature of dark matter. It is therefore essential to search for observational signatures from alternative theories of gravity beyond GR. 

As an effective four-dimensional version of five dimensional string theory, braneworld models are of particular interest among the various alternative theories of gravity. In these models, the usual physical universe nests on a three-brane embedded in a five dimensional spacetime (bulk). All of the matter fields are confined to the three-brane, and only gravity can freely propagate in both brane and bulk. Since gravity enters as an extra spatial dimension\cite{Dadhich:2000am,PoncedeLeon:2007bq,Molina:2016tkr}, studying gravitational effects in braneworld models may shed light on the physical signatures of five dimensions on the four dimensional physical world. Due to high-energy corrections and Weyl stresses from bulk gravitons, static black hole solutions localized on the brane are no longer described by the Schwarzschild metric \cite{Dadhich:2000am}. Furthermore, the five-dimensional Einstein field equations admit a significantly richer family of spherically symmetric solutions on the brane than those allowed in four-dimensional general relativity. The first brane-localized black hole solution, derived in Ref. \cite{Dadhich:2000am}, possesses the identical form to the standard Reissner–Nordstr\"{o}m metric, with a tidal Weyl parameter playing the role of the electric charge. The other black hole solutions in the braneworld
model and the corresponding observable effects have also
been widely studied\cite{PoncedeLeon:2007bq,Molina:2016tkr,Heydar-Fard:2007ahl,Randall:1999ee,Lin:2022ldu,Tadros:2022qvg,Harko:2004ui,Mak:2004hv,Harko:2005fq,Harko:2005yk,Amarilla:2011fx,Eiroa:2017uuq,Abdujabbarov:2017pfw,Banerjee:2019nnj,Hou:2021okc}. 

Horizon-scale images of M87* and Sgr A* recently released by the Event Horizon Telescope provide a unique laboratory for testing gravity in the strong-field regime \cite {EventHorizonTelescope:2022xqj,EventHorizonTelescope:2022wkp,EventHorizonTelescope:2019ths,EventHorizonTelescope:2019uob}. This is mainly attributed to that the morphologies and brightness distributions of these black hole images encode a wealth of invaluable information for gaining insights into the physics in the strong-field regions in the vicinity of black holes.
Black hole images are dominated by  the intrinsic properties of the central black hole, the surrounding matter distribution and the corresponding  radiation transfer. In the real astronomical environment, the accretion flows near black holes
acts as a major component of the emission source and play a crucial role in black hole imaging. Generally, a complete description of the dynamics of accretion flows must resort to highly accurate general relativistic magnetohydrodynamics (GRMHD) simulations \cite{EventHorizonTelescope:2022wkp,Banerjee:2019nnj}. However, the associated high computational cost  poses enormous challenges for systematically exploring the vast parameter space. For the current EHT targets like M87* and Sgr A* operating at low accretion rates, the accretion flow is  radiatively inefficient  and the corresponding disk is typically hot, geometrically thick and optically thin at an observing frequency of 230 GHz. Furthermore, a semi-analytic  model is confirmed to be a feasible alternative for describing the radiatively inefficient accretion flows (RIAFs) near these supermassive black holes and characterizing their observational signatures \cite{Yuan:2003dc,Pu:2018ute}. With this semi-analytic RIAFs model, the corresponding horizon-scale images for Kerr and Kerr-Sen black hole have been analyzed along with the influences of black hole parameters and accretion disk thickness on black hole images \cite{Jiang:2023img,Pu:2018ute,Yan:2025mlg,Pu:2016qak,Yin:2025rao}.
Given that braneworld black holes are associated with extra dimensions, it is of great significance to investigate images of braneworld black holes within the RIAF framework and to probe the corresponding combined effects of extra dimensions and accretion disk thickness. Here, we consider a spherically symmetric solution in
the brane world obtained by Casadio, Fabbri and Mazzacurati
\cite{Casadio:2001jg}. The properties of thin accretion disks \cite{Pun:2008ua}  and thick accretion disk configurations  \cite{Wei:2023tiy}around the
Casadio–Fabbri–Mazzacurati (CFM) black hole have also been investigated, which could offer the possibility to directly test physical models with extra
dimension using astrophysical observations from accretion disks. 

The paper is organized as follows. In section. II, we briefly
introduce the CFM braneworld black hole and the RIAF model. 
In section. III, we perform  general relativistic radiative transfer simulations and generate synthetic images consistent with Event Horizon Telescope observations of M87* for CFM braneworld black holes. Then, we also probe effects of the tidal parameter on black hole images and analyze the possibility of detecting tidal parameter from black hole image using future ngEHT and BHEX projects. In Section. IV,  we end the paper with a summary.

\section{Radiatively Inefficient Accretion Flow around Braneworld Black Holes}
In this section, we briefly review a static and spherically symmetric braneworld black hole solution, known as the Casadio-Fabbri-Mazzacurati (CFM) black hole. Starting from the five-dimensional vacuum Einstein equations and projecting them onto a four-dimensional brane, one can obtain the induced field equations\cite{Bronnikov:2003gx,Israel:1966rt}
\begin{eqnarray}
    R^{(4)}_{\mu\nu}=\Lambda_4g^{(4)}_{\mu\nu}-E^{(4)}_{\mu\nu},\label{effEfield}
\end{eqnarray}
where $R^{(4)}_{\mu\nu}$ and $g^{(4)}_{\mu\nu}$ denote the Ricci tensor and metric on the brane, and $\Lambda_4$ represents the effective four-dimensional cosmological constant. The traceless tensor $E_{\mu\nu}$ describes nonlocal gravitational effects originating from the bulk, which can be regarded as a tidal imprint of the higher dimensional bulk geometry on the brane. The induced field equations (\ref{effEfield}) admit a static, spherically symmetric solution \cite{Casadio:2001jg}
\begin{eqnarray}\label{metric}
ds^{2}
&=& \left(1 - \frac{2M}{r}\right) dt^{2}
- \frac{1 - \dfrac{3M}{2r}}
{\left(1 - \dfrac{2M}{r}\right)\left(1 - \dfrac{\gamma M}{2r}\right)} dr^{2}
- r^{2} d\theta^{2}-r^2\sin^2\theta d\phi^{2},
\end{eqnarray}
where $M$ represents the mass parameter and $\gamma$ is a dimensionless quantity that characterizes the tidal effect of the bulk geometry, with its magnitude influencing the brane spacetime properties. As $\gamma=3$, the metric reduces exactly to the
Schwarzschild solution. In general, the geometry corresponds
to a black hole with a single event horizon when $\gamma<4$,  becomes extremal with coinciding horizons as $\gamma$ tends to the critical value $\gamma=4$ and transitions to a traversable wormhole geometry $\gamma>4$. 
In the black hole regime with $\gamma<4$, the black hole horizon is located at $r_h=2M$, identical to that of the Schwarzschild black hole and independent of the value of $\gamma$. However,  the  Hawking temperature is given by\cite{Liu:2007qwa}
\begin{equation}
T_{\mathrm{BH}}
= \frac{1}{8\pi M}
\sqrt{1 - \frac{3(\gamma - 3)}{2}},
\end{equation}
which explicitly depends on the tidal parameter $\gamma$. The tidal parameter $\gamma$ is related to the only non-vanishing parameterized post-Newtonian (PPN) parameter $\beta$ by $\gamma=4\beta-1$. Agreement with observations from solar system measurements requires $|\beta-1|\ll 1$, which means that $\frac{5}{3}<\gamma<\frac{13}{3}$ \cite{Casadio:2001jg}. To ensure that  the Hawking temperature is real, one has $\gamma<\frac{11}{3}$. Therefore, in this work, we focus on only the cases with $\frac{5}{3}<\gamma<\frac{11}{3}$. 

Black hole images are jointly determined by both the black hole parameters and  the properties of the surrounding plasma \cite{Younsi:2021dxe,Ozel:2021ayr}. 
 For the current observed targets of the EHT including M87* and Sgr A*, their surrounding hot plasmas are considered to be part of a RIAF. Here, we adopt a semi-analytical optically thin RIAF model \cite{Yuan:2003dc,Pu:2018ute} around the braneworld black hole to perform general relativistic ray-tracing numerical codes for simulating the black hole images\cite{Noble:2007zx,Moscibrodzka:2017lcu}. 
Within the framework of the RIAF model, the number density and temperature of electrons can be
    written as a hybrid combination of radial and exponential functions given by\cite{Saurabh:2025kwb}
\begin{equation}
n_e
= n_{e,0}\, (\frac{r}{r_g})^{-\delta}
\exp\!\left[
-\frac{1}{2}\left(H \tan\theta\right)^{-2}
\right] ,
\end{equation}
\begin{equation}
T_e
= T_{e,0}\, (\frac{r}{r_g})^{-\alpha} ,
\end{equation}
where $r_g=\frac{GM}{c^2}$ is a gravitational radius and $H$ denotes the accretion disk thickness. $n_{e,0}$ and $T_{e,0}$ are the normalization parameters for the electron number density and temperature, respectively,  and $\delta$ and $\alpha$ are the corresponding power-law indices. The temperature index $\alpha$ is observationally constrained by the spectral energy distribution \cite{Pu:2018ute} and VLBI brightness temperature measurements \cite{Pu:2016qak}. The magnetic field strength
of the toroidal field is then given by
\begin{equation}
\frac{B^{2}}{8 \pi}=\frac{1}{10} n_{e} \frac{m_{\mathrm{p}} c^{2} r_{g}}{6 r},
\end{equation}
where $m_p$ is the mass of proton\cite{Saurabh:2023otl}.
Following Ref. \cite{Pu:2016qak,Broderick:2021ohx} the four-velocity of the accretion flow can be constructed by interpolating between Keplerian rotation and geodesic free-fall motion, and its form can be written as\cite{Saurabh:2025kwb}
\begin{eqnarray}
u^\mu = (u^t, u^r, 0, \Omega u^t),
\label{eq:four-velocity}
\end{eqnarray}
with
\begin{eqnarray}
u^r = u_K^r + \kappa_{ff} (u_{ff}^r - u_K^r),\quad\quad\quad \Omega = \Omega_K + (1 - \kappa_{K})(\Omega_{ff} - \Omega_K).
\end{eqnarray}
Here $u^r_K$ ($\Omega_K$) and $u^r_{ff}$ ($\Omega_{ff}$) correspond to the radial (angular) velocities of  the flows moving along Keplerian motion and free fall motion, respectively. The coefficient $\kappa$ is a regulating
parameter. As $(\kappa_{ ff},\kappa_{ K}) = (0,1)$ and $(\kappa_{ ff},\kappa_{ K}) = (1,0)$, the motion of flows corresponds to pure Keplerian motion and free fall motion, respectively. A hybrid motion model is better suited to describe the actual astrophysical flow in $M87*$, in which a sub-Keplerian azimuthal velocity profile and advective effects are anticipated\cite{Takahashi:2011dr}. 
Here, we set to $(\kappa_{\rm ff},\kappa_{\rm K}) = (0.5,0.5)$ in the fiducial RIAF model. Finally, the velocity component $u_t$ of flows can be derived by enforcing the normalization condition of the four-velocity $u_{\mu}u^{\mu}=-1$.

 Since the non-thermal emission is less dominant for the 230 GHz waveband, we here model the synchrotron emissivity in the accretion disk using a relativistic thermal (Maxwell–Jüttner) electron distribution.
 To ensure that the observed flux of thermal synchrotron radiation matches the measured value from M87*, the normalized electron number density $n_{e,0}$ and the electron temperature $T_{e,0}$ are set to $n_{e,0} \approx 10^6 \rm cm^{-3}$ and  $T_{e,0} \approx 10^{11} \rm K$, respectively \cite{Pu:2018ute}. The remaining fiducial parameters are listed in Table I.
\begin{table}[ht]
\centering
\caption{}
\label{tab:parameters}
\begin{tabular}{lcl}
\hline
\textbf{Parameter} & \textbf{Value} & \textbf{Parameter Description} \\
\hline
$M_{\text{BH}}$ & $6.4 \times 10^9 \, M_\odot$ & Black hole mass \\
$D_s$ & $16.9 \times 10^6 \, \text{pc}$ & Distance to the source \\
$\delta$ & $1.1$ & $n_e$ power law index \\
$\alpha
$ & $0.84$ & $T_e$ power law index \\
$\kappa_K$ & $0.5$ & Keplerian parameter \\
$\kappa_{\text{ff}}$ & $0.5$ & Radial infall parameter \\
$n_{e,0}$ & $\approx 10^6 \, \text{cm}^{-3}$ & Number density \\
$T_{e,0}$ & $\approx 10^{11} \, \text{K}$ & Temperature of electrons \\
$\nu_{\text{obs}}$ & $230 \, \text{GHz}$ & Observing frequency \\
FOV & $200 \, \mu\text{as}$ & Field of view \\
\hline
\end{tabular}
\end{table}

\section{Images of Braneworld black holes with radiatively inefficient accretion flows}
\begin{figure}
\includegraphics[width=\textwidth]{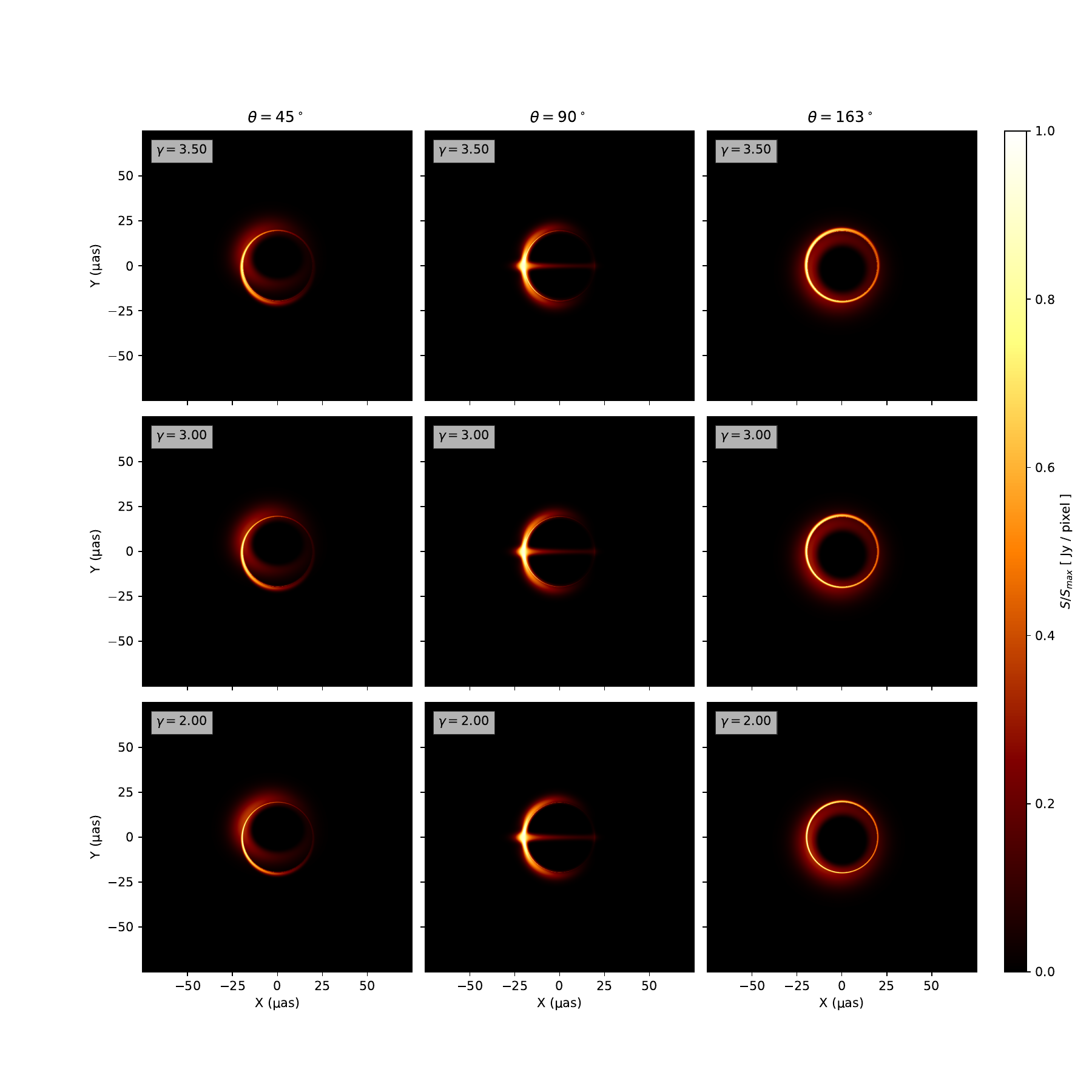} 
\caption{Simulated images of  RIAF around braneworld black holes for different tidal parameters $\gamma$ and  inclination angles $\theta$. From left to right, the inclination angles correspond to $\theta=45^\circ, 90^\circ$ and $163^\circ$, while from top to bottom the tidal parameter takes values $\gamma=3.5$, $3$ and $2$. Here we set observed frequency  $\nu_{obs}=230 GHz$, accretion disk thickness $h=0.1$ and mass of black hole $M_{BH}=6.4\times10^9M_{\bigodot}$. Additionally, all images have identical radiative flux value 0.5 Jy.}
\label{imagesth}
\end{figure}

\begin{figure}
\centering
\includegraphics[width=2.32in]{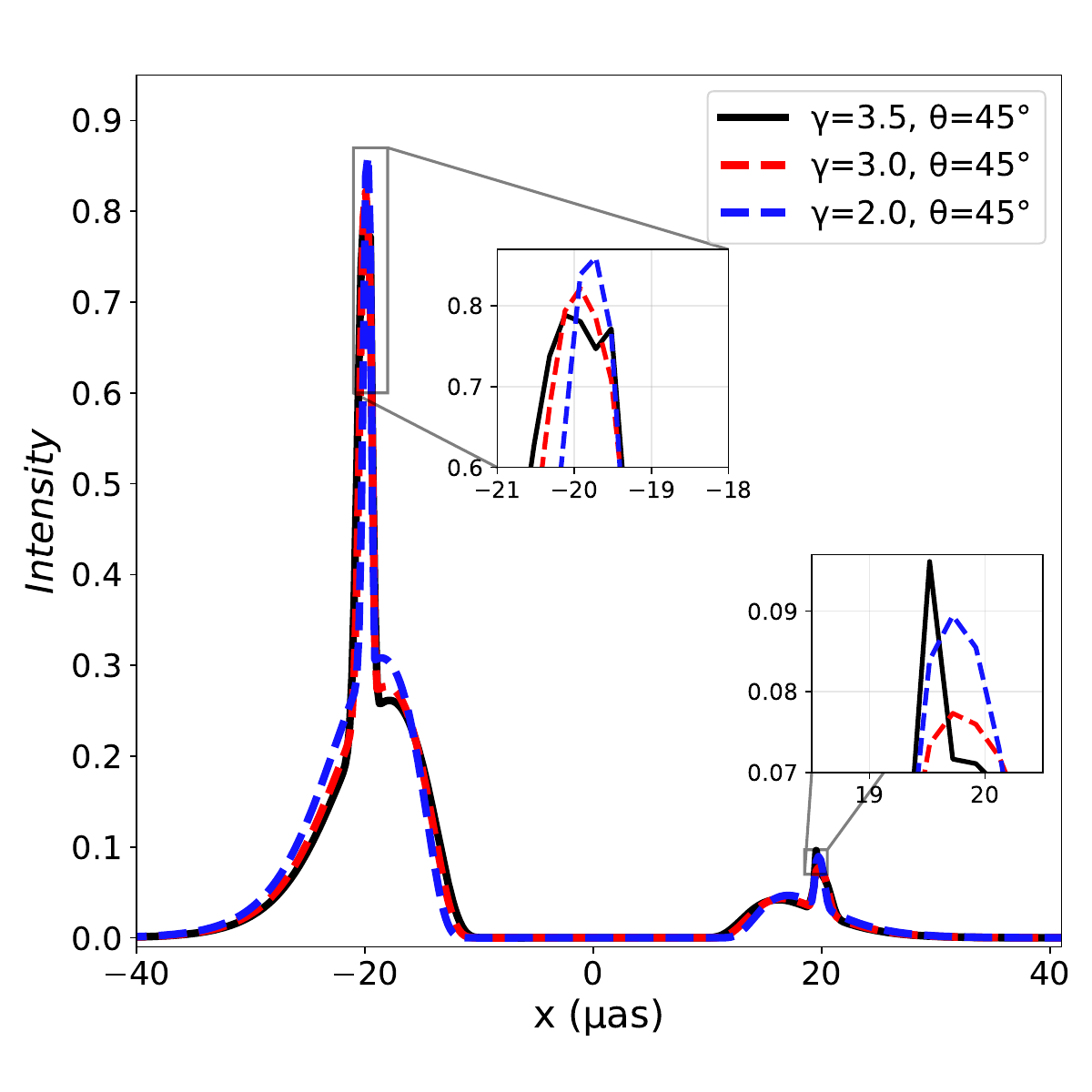} \includegraphics[width=2.32in]{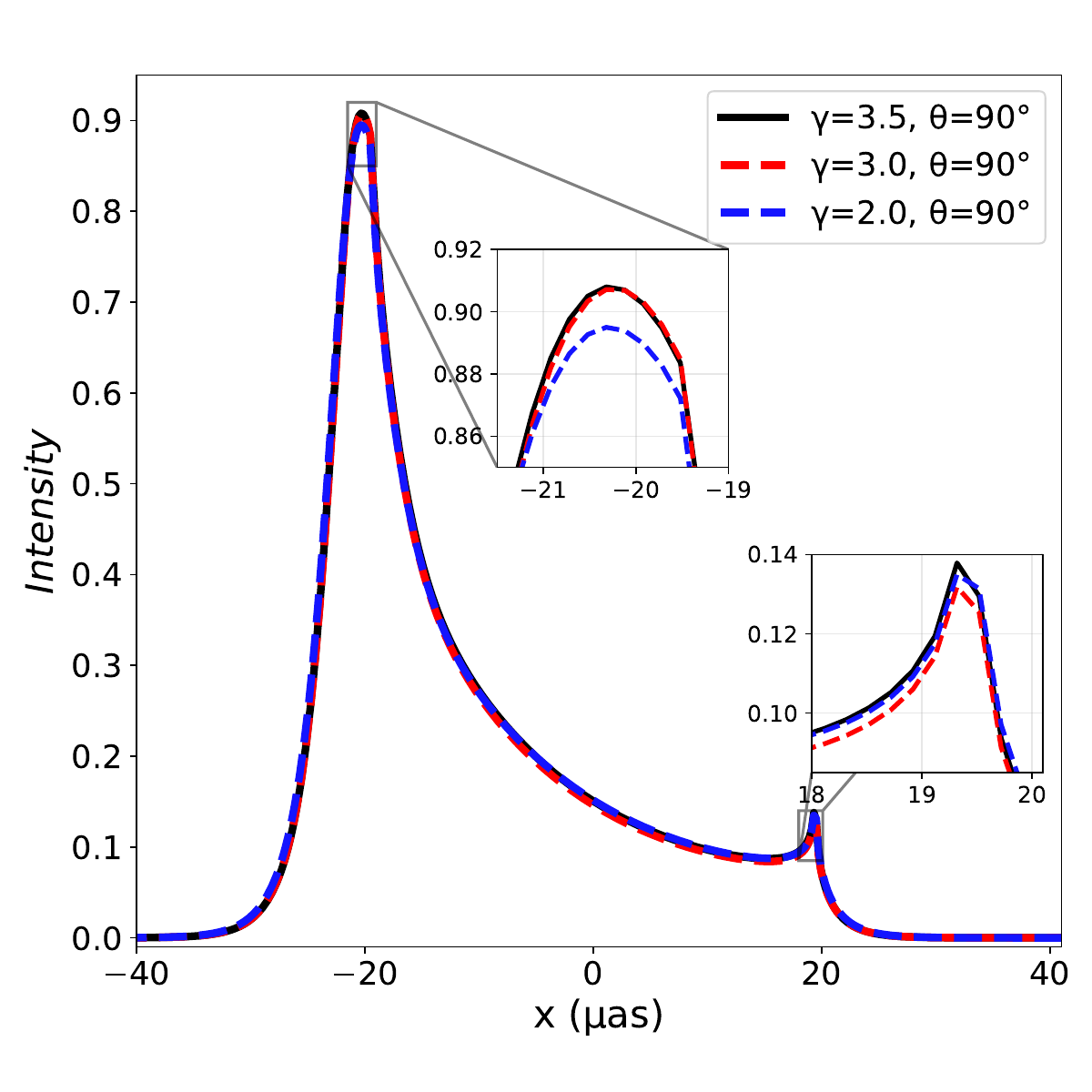}  
\includegraphics[width=2.32in]{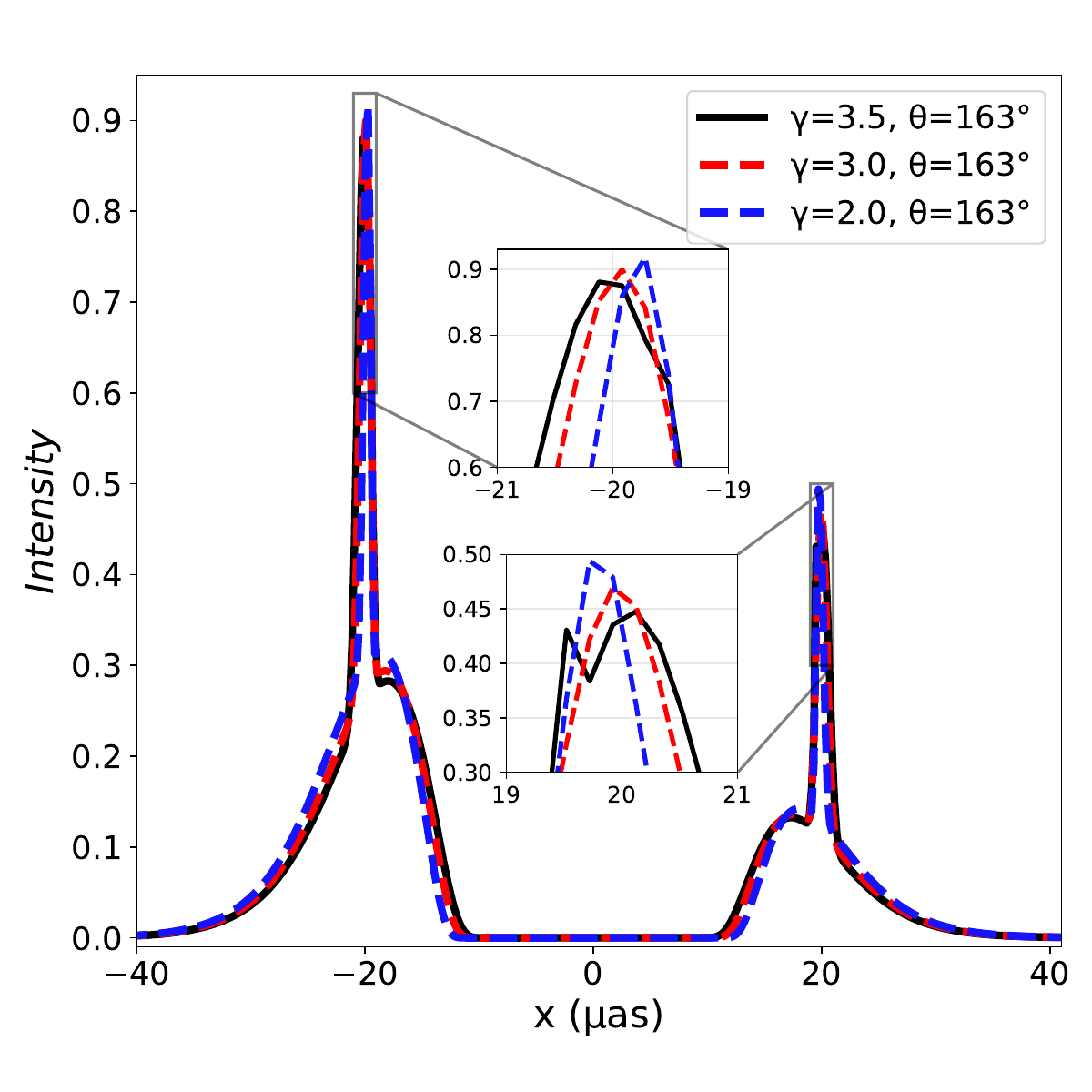} \includegraphics[width=2.33in]{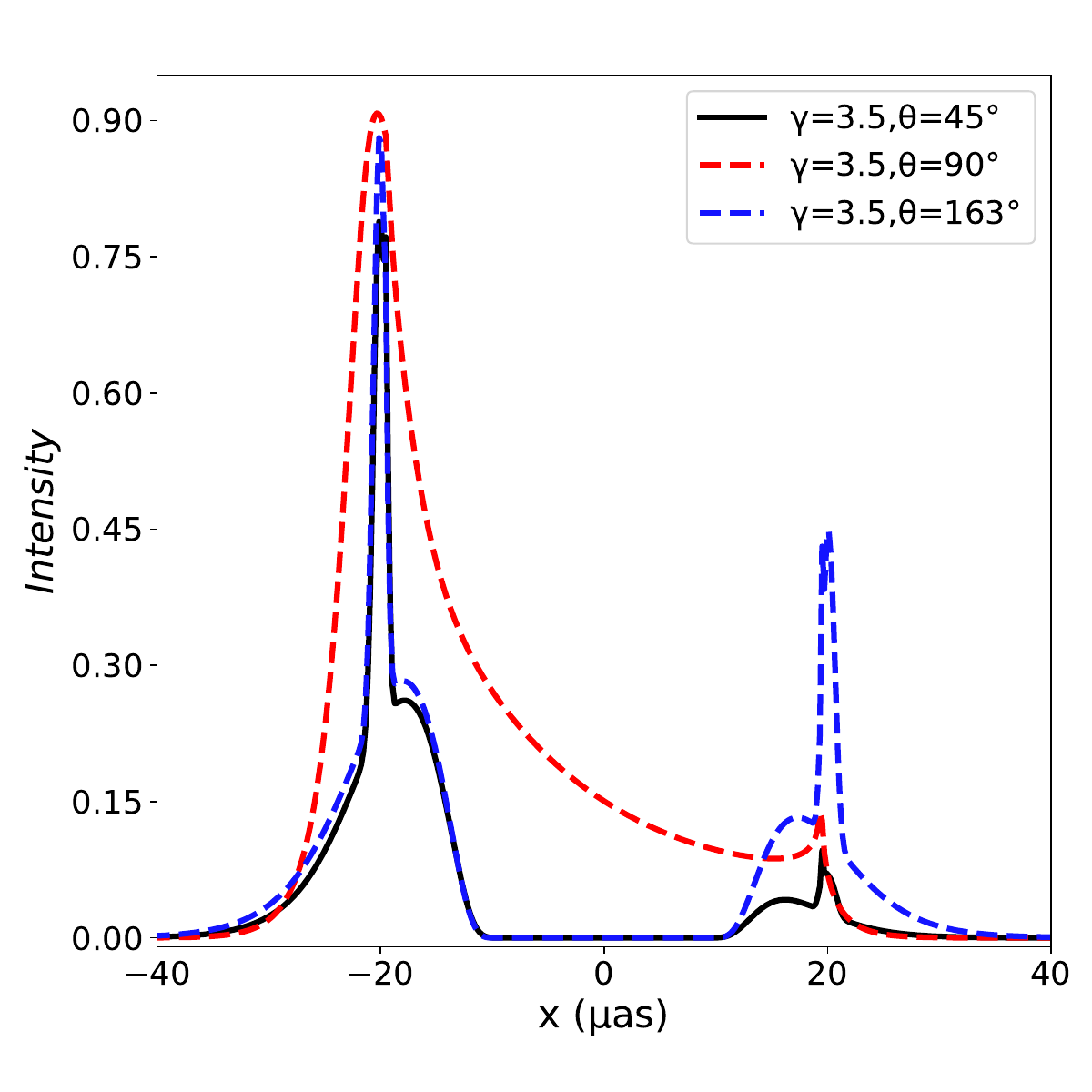} \includegraphics[width=2.32in]{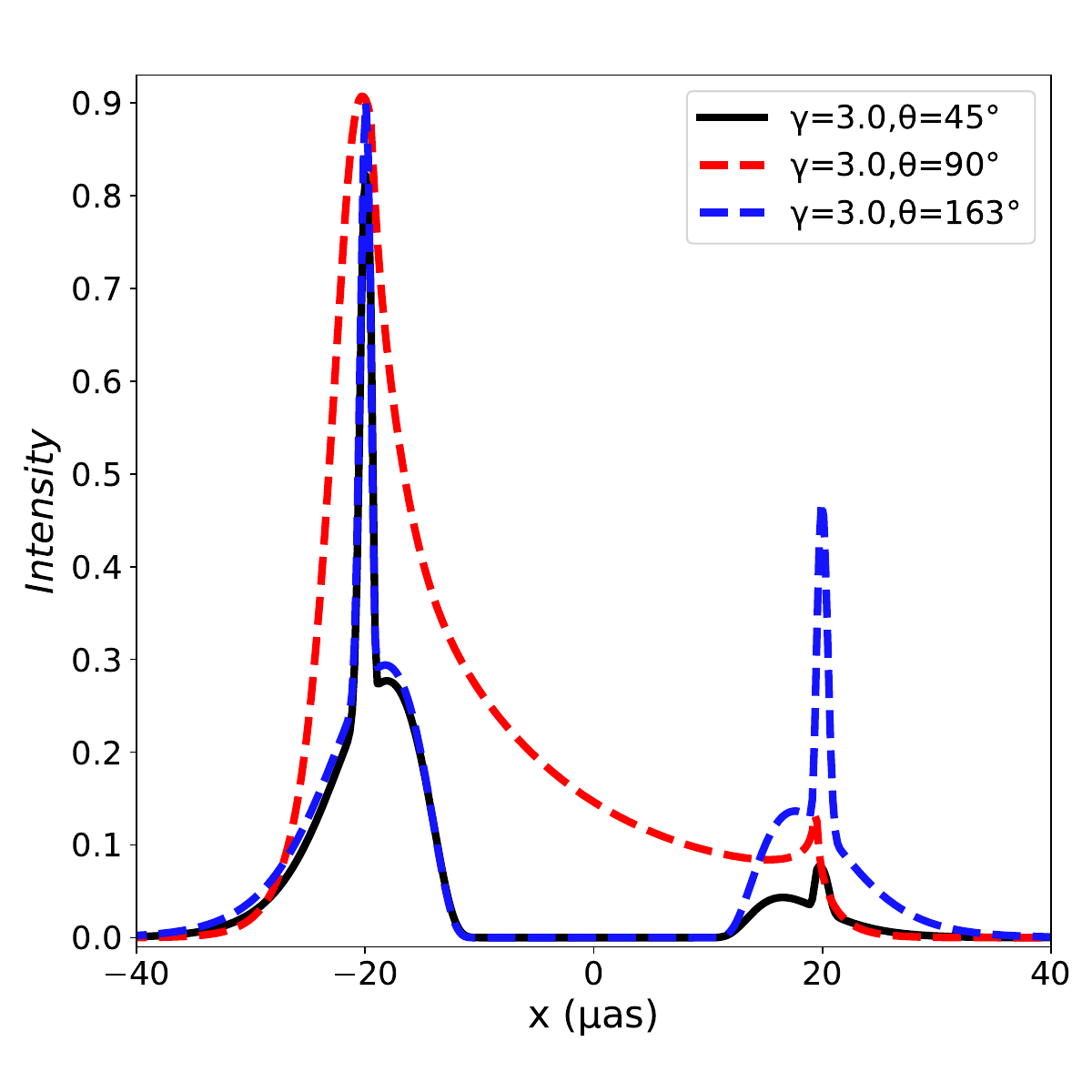}  
\includegraphics[width=2.32in]{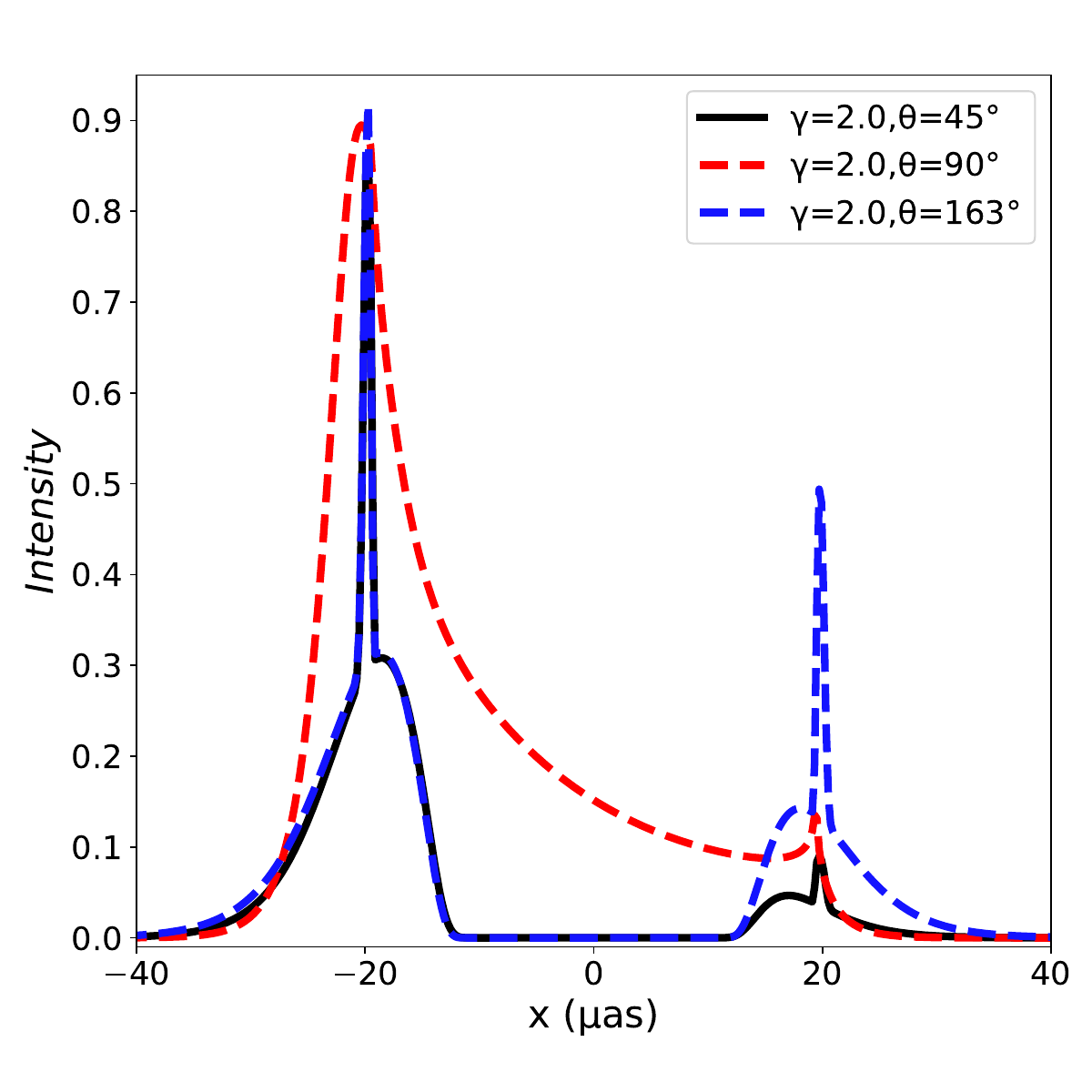} 
\caption{The normalized intensity profiles along the horizontal cross section $y=0$ for the model images of Fig.\ref{imagesth} for different tidal parameters$\gamma$ and inclination angles $\theta$. }
\label{imagesiis}
\end{figure}

Lets now to probe effects of the tidal parameter $\gamma$ on images of braneworld black holes with RIAFs. In Fig.\ref{imagesth}, we present the simulated  braneworld black hole images for different tidal parameter $\gamma$ and inclination angle $\theta$. Similar to other black hole systems, the images of braneworld black holes exhibit a bright ring surrounding the central dark region.
The asymmetric bright ring and the pronounced brightness enhancement on the side approaching the observer are also characteristic features of braneworld black hole images. 
Moreover, we find the effects of the tidal parameter $\gamma$ is insignificant on the black hole images. In Fig.\ref{imagesiis}, we present the normalized intensity profiles along the horizontal cross section $y=0$ of the images in Fig.\ref{imagesth}. The variation of the peak value and peak width of the luminosity distribution with the tidal parameter $\gamma$ depends on the inclination angle $\theta$. The primary peak value in the left decreases with the parameter $\gamma$ for $\theta=45^{\circ}$ and $\theta=163^{\circ}$, but slightly increases for $\theta=90^{\circ}$. The secondary peak value in the right decreases with the parameter $\gamma$ for $\theta=163^{\circ}$, while for $\theta=45^{\circ}$ and $\theta=90^{\circ}$, it first decreases and then increases. The peak width generally increases with the parameter $\gamma$, except for the secondary peak at $\theta=45^{\circ}$, where it shows a decreasing trend. For a fixed tidal parameter $\gamma$,  the primary peak value exhibits a non-monotonic trend, first increasing and then decreasing with 
$\theta$, whereas the secondary peak value increases monotonically with  $\theta$.
Additional, we find that the inner shadow diameter decreases with the parameter $\gamma$ for a fixed inclination $\theta$, but is approximately equal at inclination angles of $45^{\circ}$ and $163^{\circ}$ for a fixed $\gamma$.

\begin{figure}
\centering
\includegraphics[width=\textwidth]{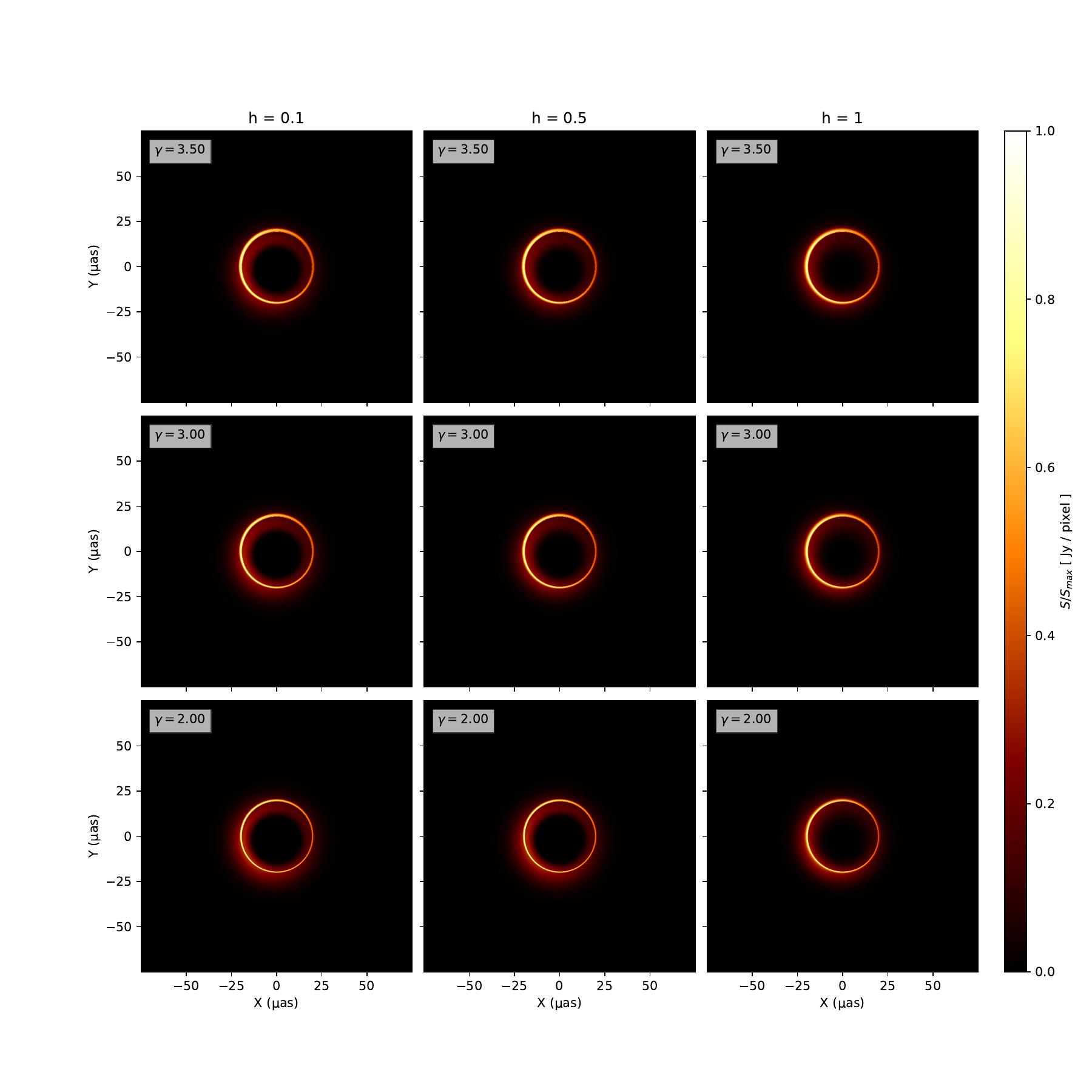} 
\caption{Simulated images of RIAF around braneworld black holes with different tidal parameters $\gamma$ and accretion disk thicknesses $h$. The accretion disk thicknesses in the left and right panels are set to $ h=0.1$ , $h=0.5$,and $h=1$,respectively. Here set to the observed frequency  $\nu_{obs}=230 GHz$ and inclination angle $\theta=163^{\circ}$ and mass of black hole $M_{BH}=6.4\times10^9M_{\bigodot}$. All images are set to have identical radiative flux value 0.5 Jy.}
\label{imagesh}
\end{figure}

\begin{figure}
\centering
\includegraphics[width=2.32in]{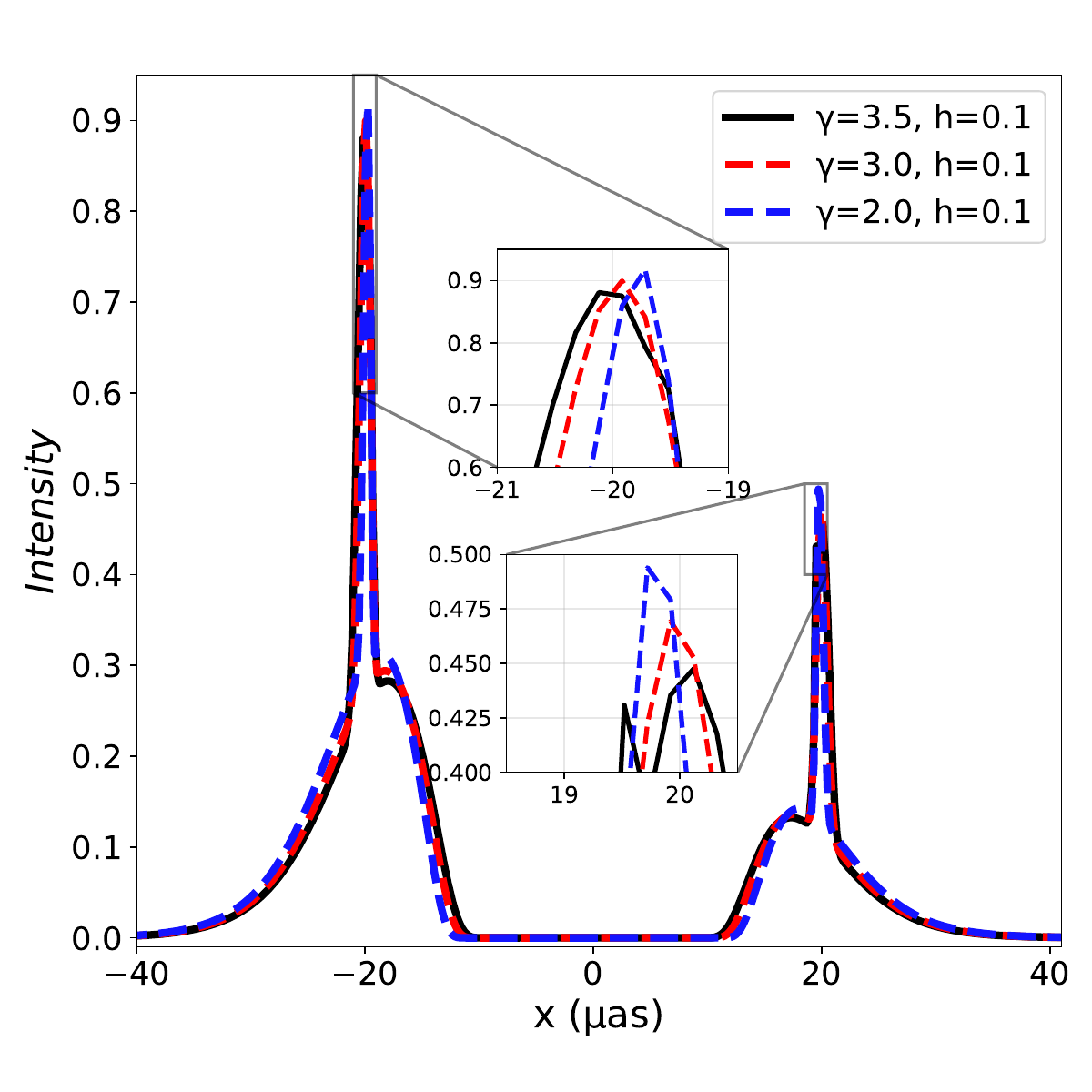} \includegraphics[width=2.32in]{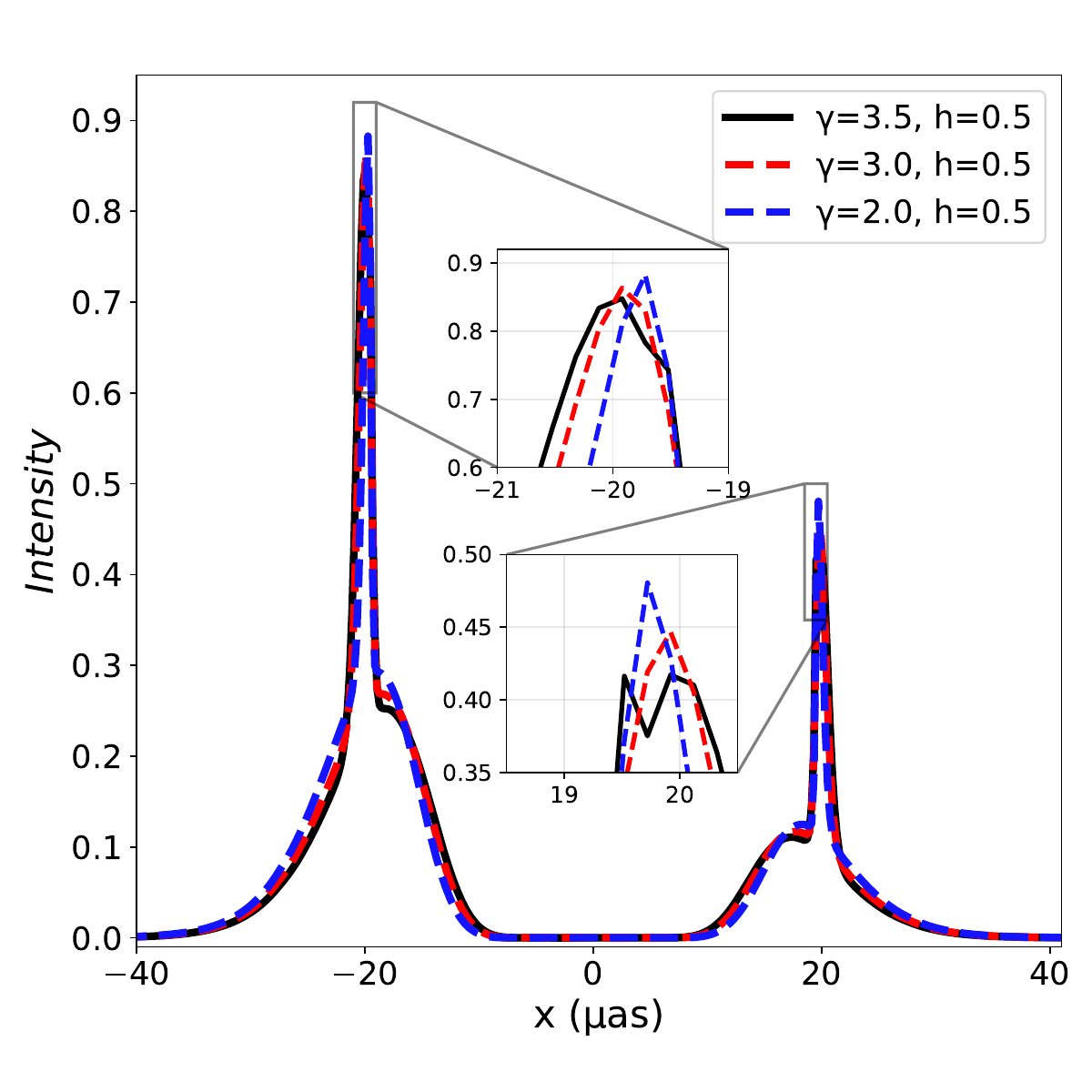}  
\includegraphics[width=2.32in]{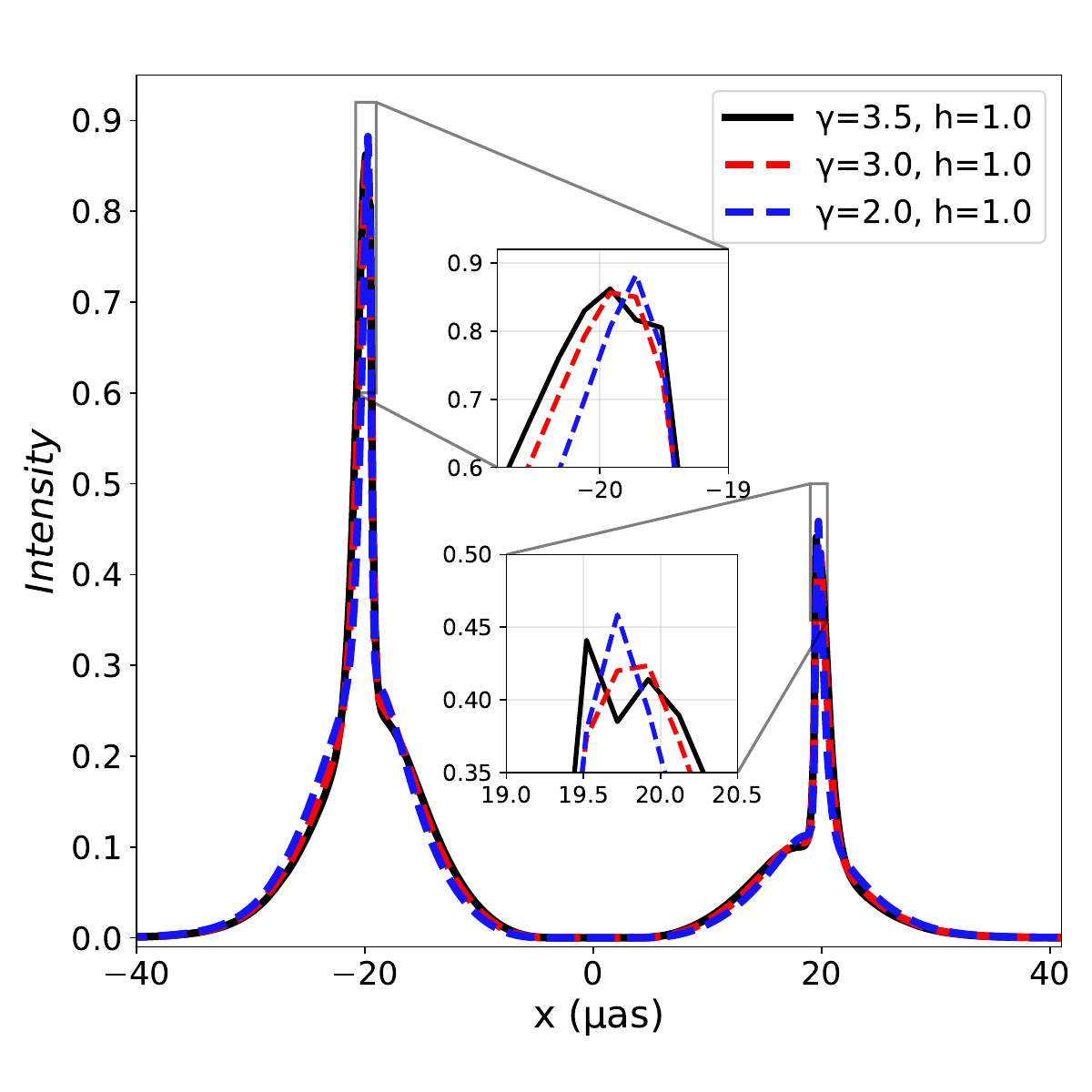} \includegraphics[width=2.33in]{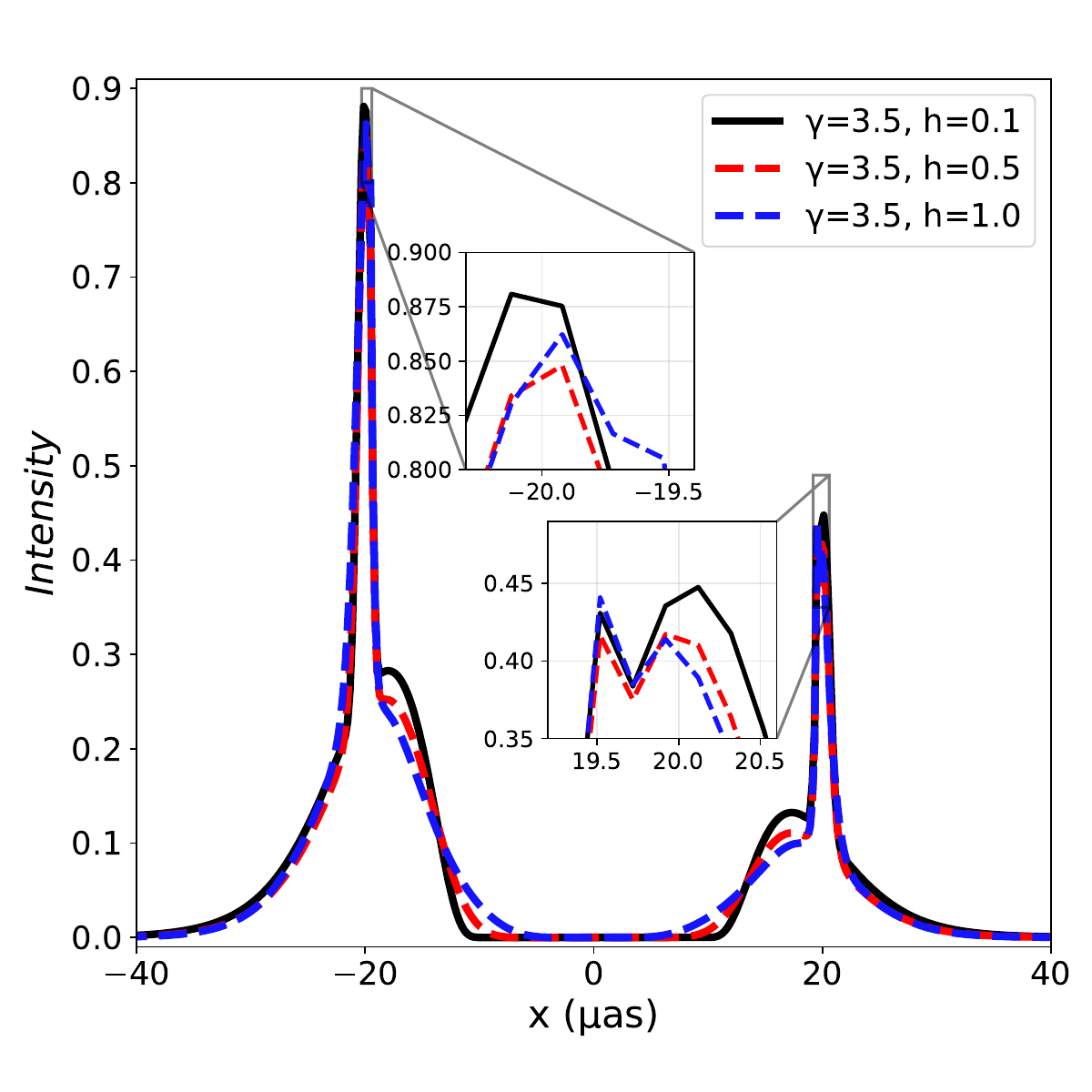} \includegraphics[width=2.32in]{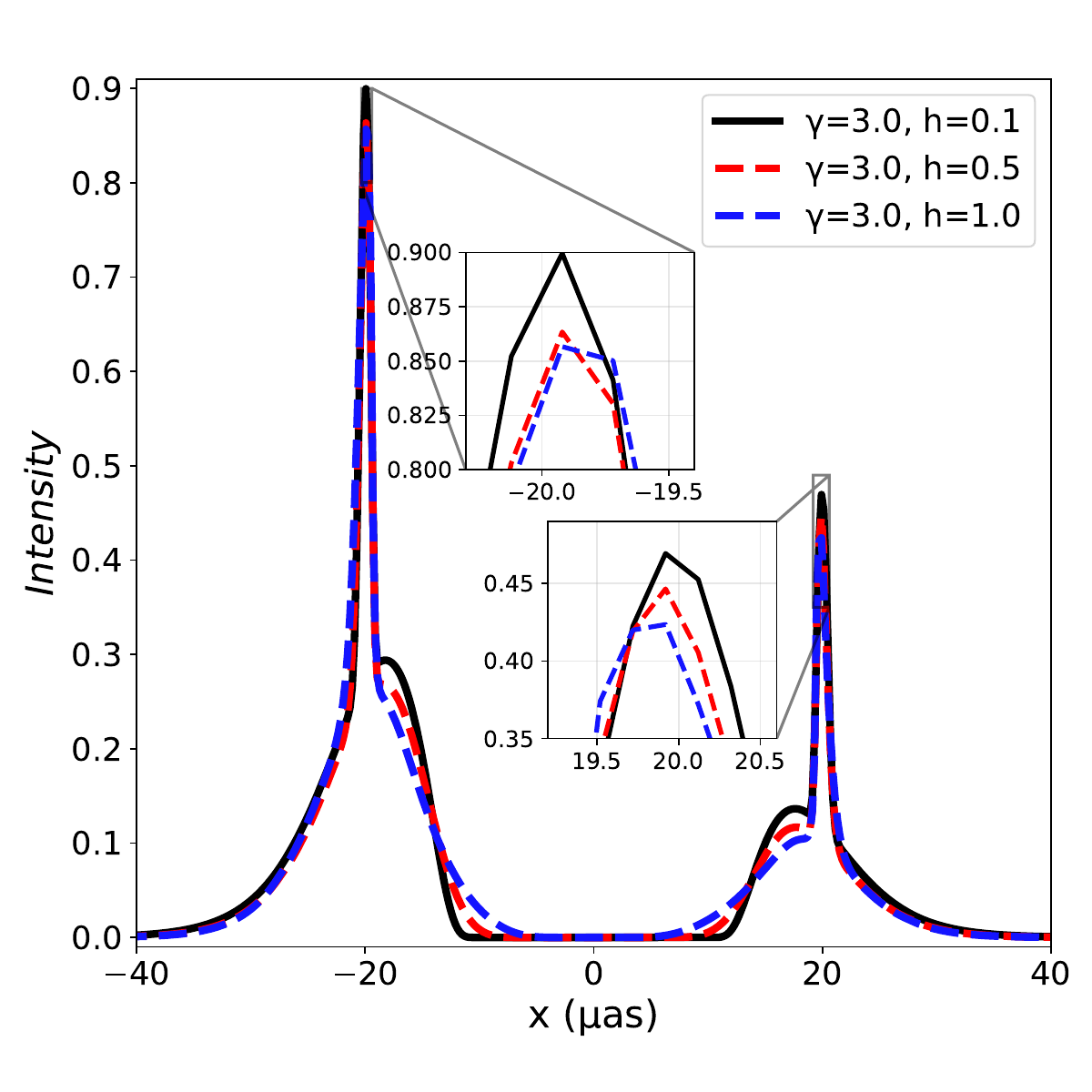}  
\includegraphics[width=2.32in]{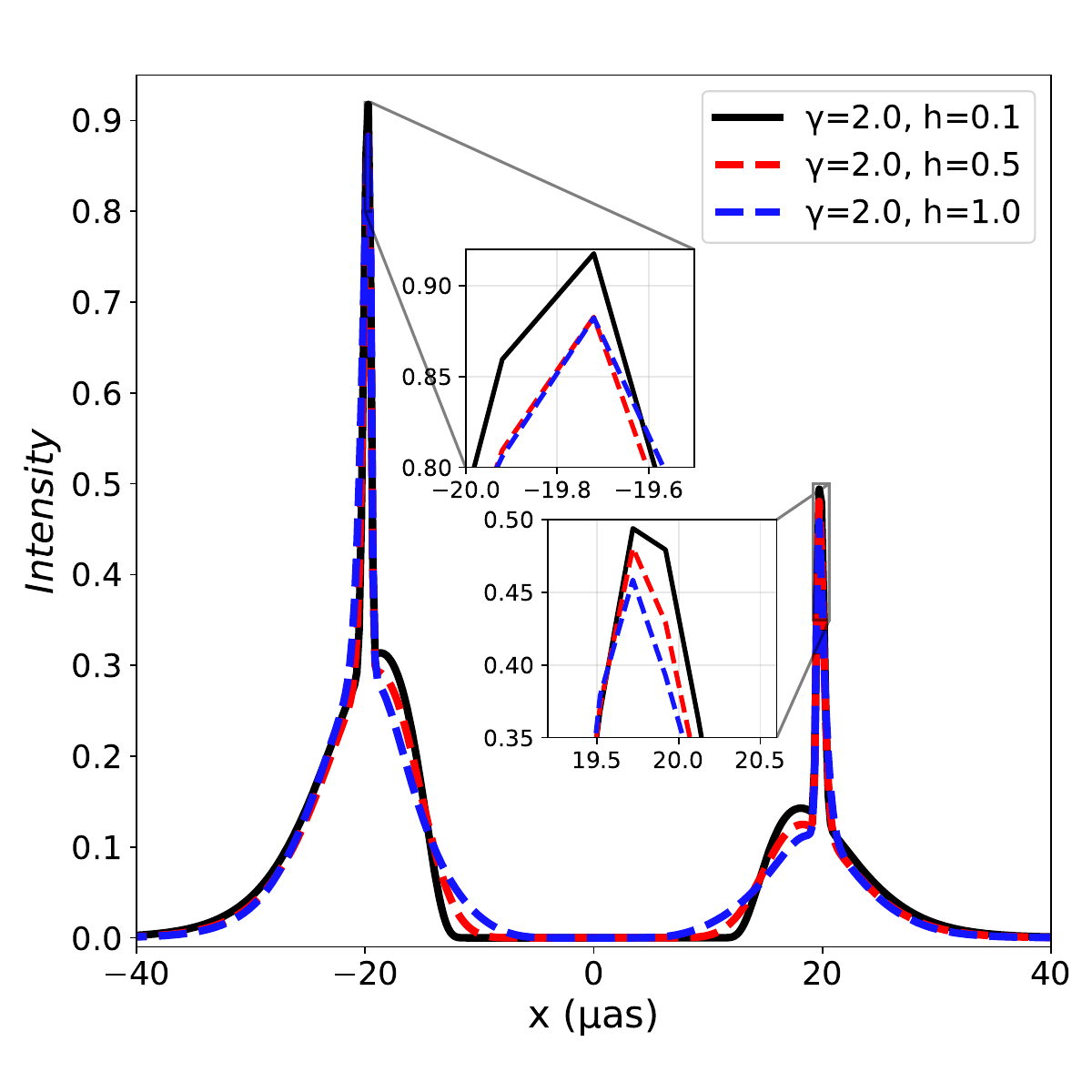}
\caption{The normalized intensity profiles along the horizontal cross section $y=0$ for the model images of Fig. \ref{imagesh} for different tidal parameters$\gamma$ and accretion disk thicknesses $h$.}
\label{imagesii01}
\end{figure}

Figs.\ref{imagesh} and \ref{imagesii01} illustrate the simulated images of braneworld black holes surrounded by RIAF  and their normalized intensity profiles along the horizontal cross section $y=0$ for different tidal parameter $\gamma$ and disk thickness $h$  for a fixed $\theta=163^{\circ}$. With the increase of $\gamma$, the primary and secondary peak values and their width decrease for different disk thickness $h$. With the increase of $h$, both the intension peak value  and the inner shadow radius decrease.

\begin{figure}
\centering
\includegraphics[width=\textwidth]{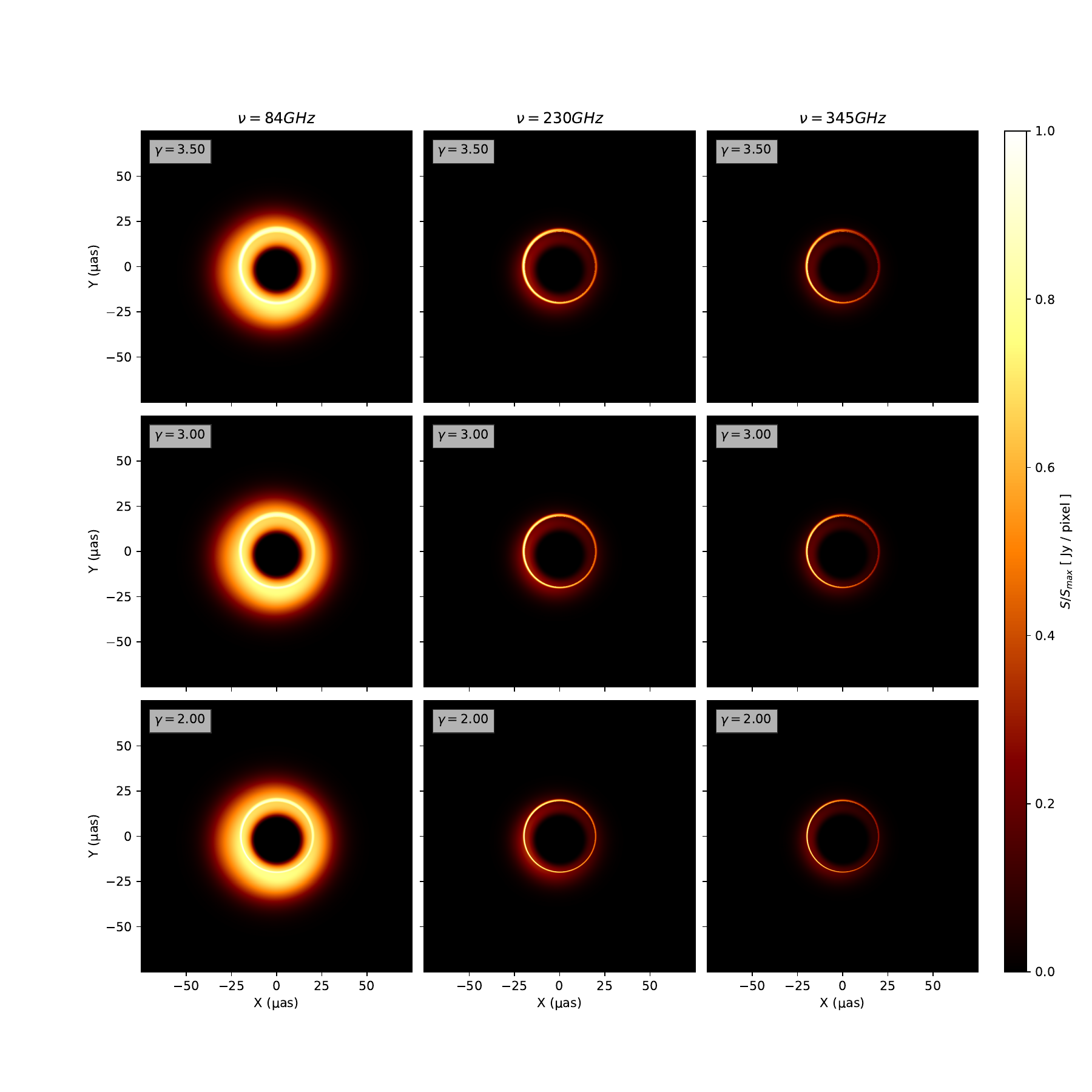} 
\caption{Simulated images of RIAF around braneworld black holes with different tidal parameters $\gamma$ and observed frequencies $\nu$ . The observed frequencies $\nu$  in the left and right panels are set to $ \nu=84GHz$ , $\nu=230GHz$,and $\nu=345GHz$,respectively. Here set the inclination angle $\theta=163^{\circ}$,  accretion disk thickness $h=0.1$, mass of black hole $M_{BH}=6.4\times10^9M_{\bigodot}$ and all images have identical radiative flux value 0.5 Jy.}
\label{imagesv}
\end{figure}
\begin{figure}
\centering
\includegraphics[width=2.32in]{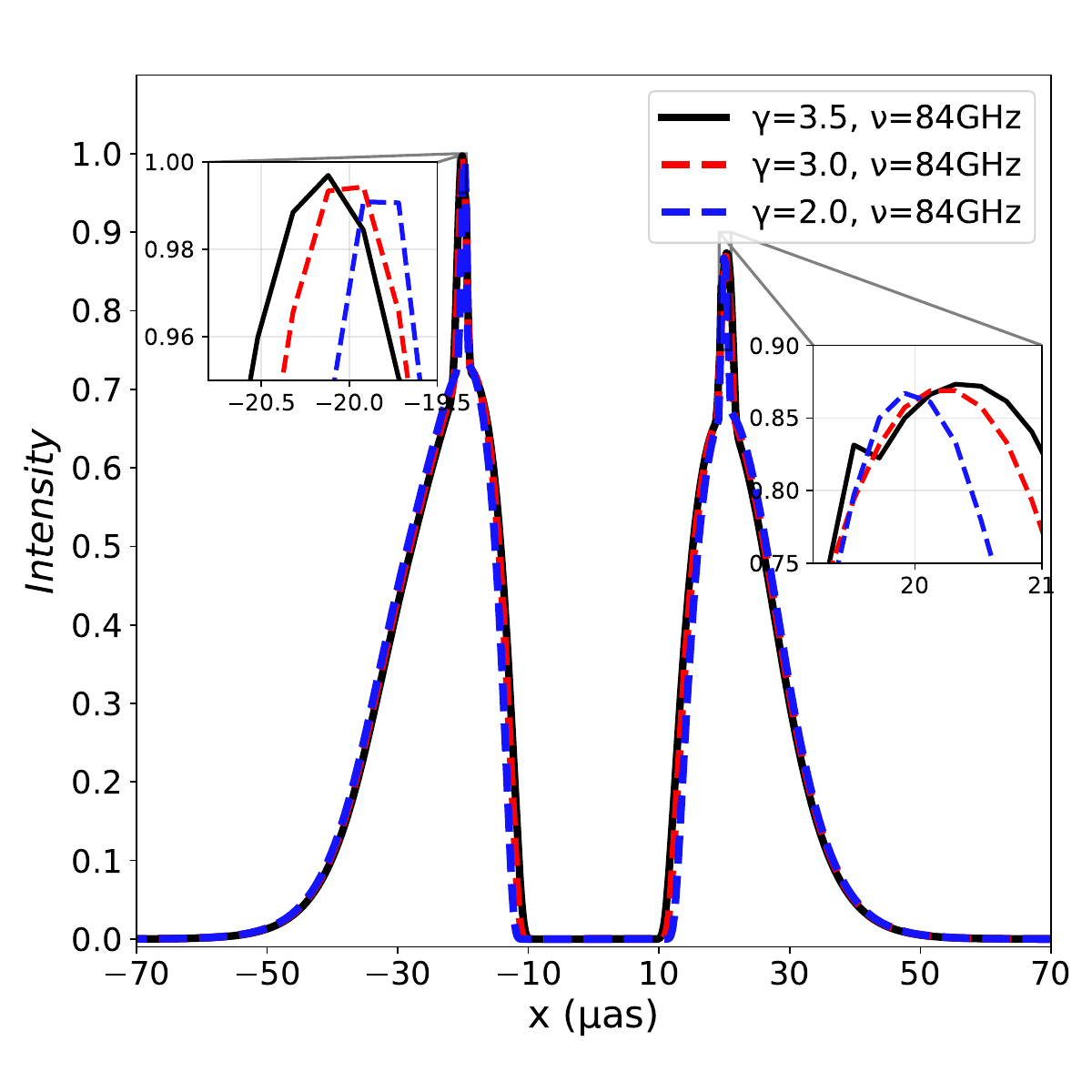} \includegraphics[width=2.32in]{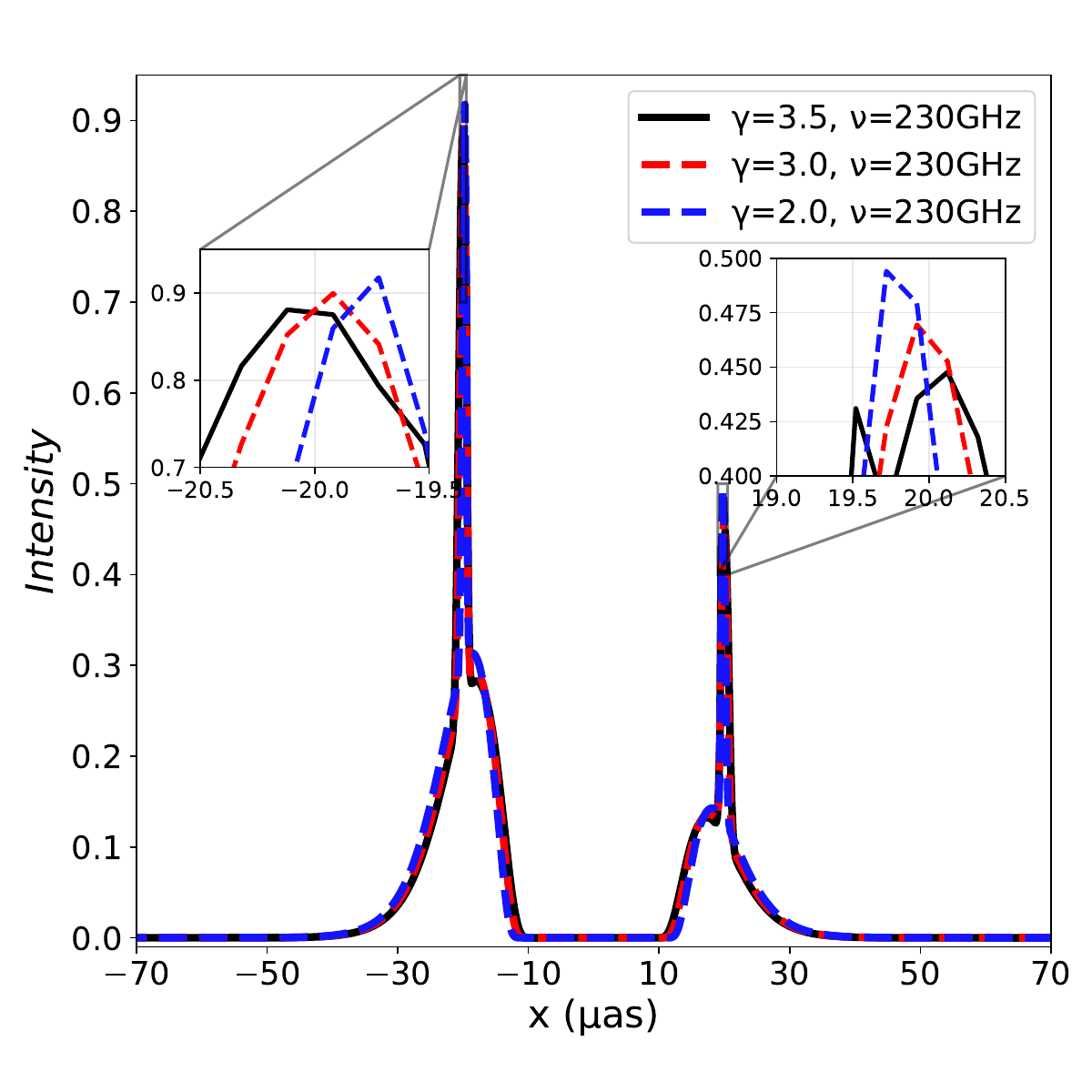}  
\includegraphics[width=2.32in]{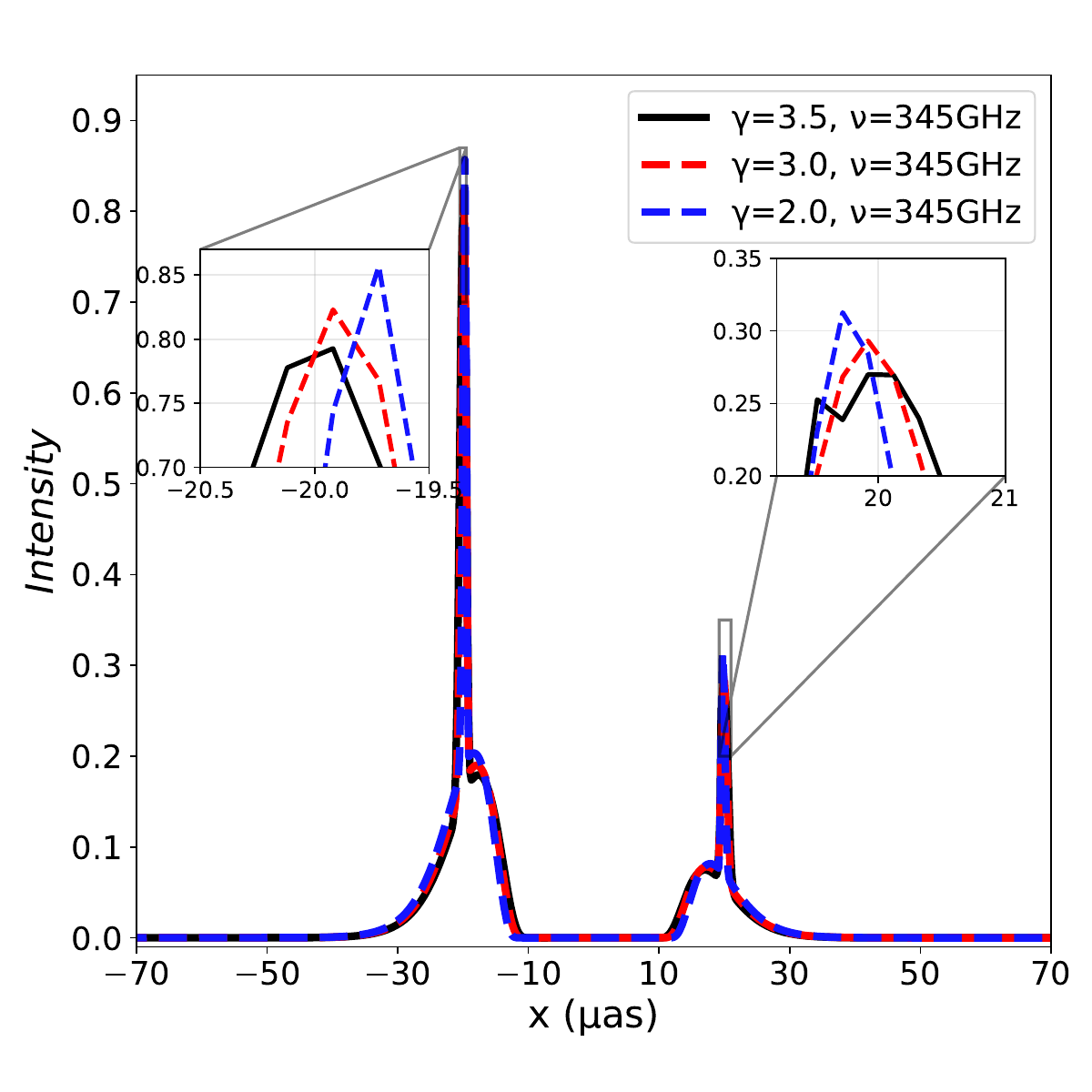} \includegraphics[width=2.32in]{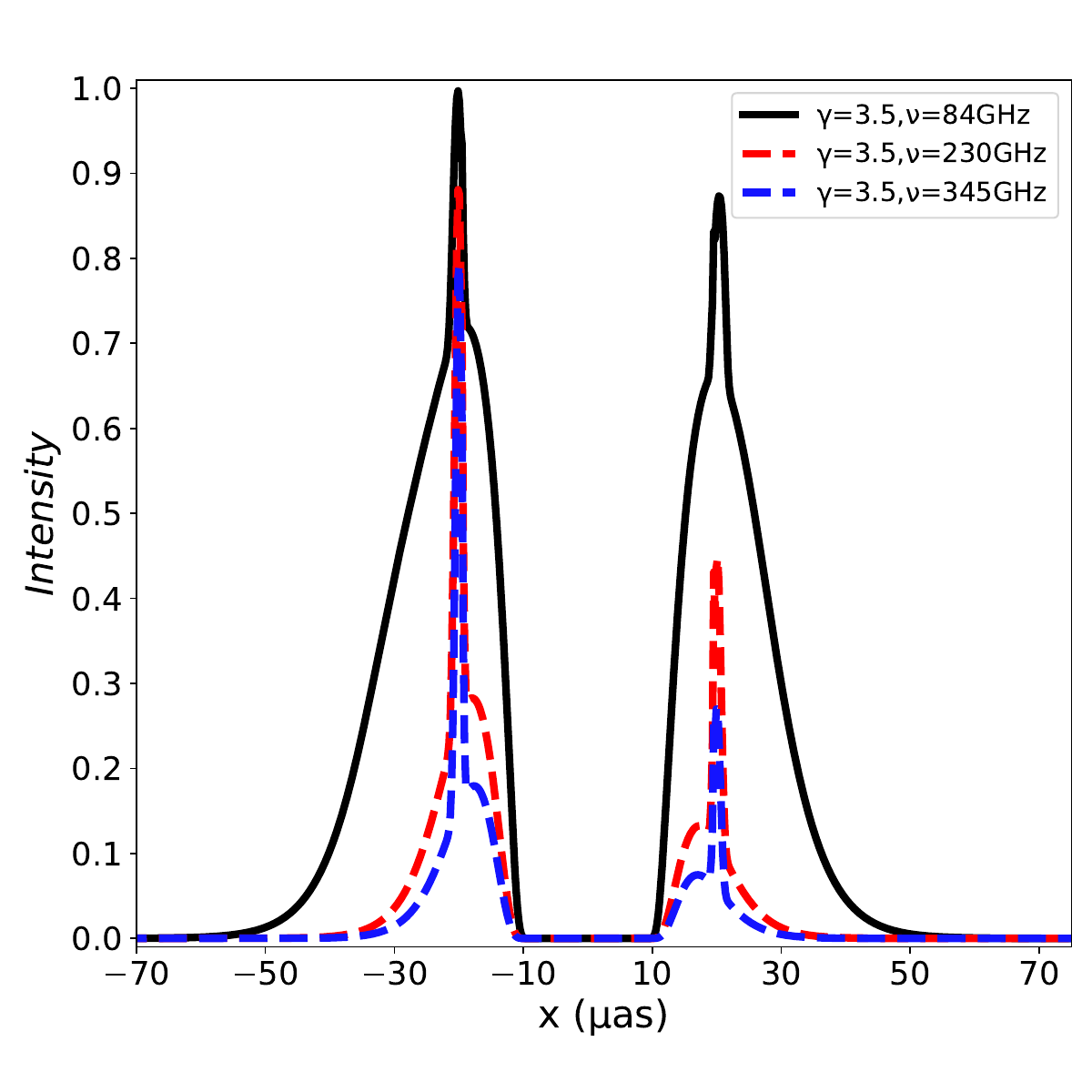} \includegraphics[width=2.32in]{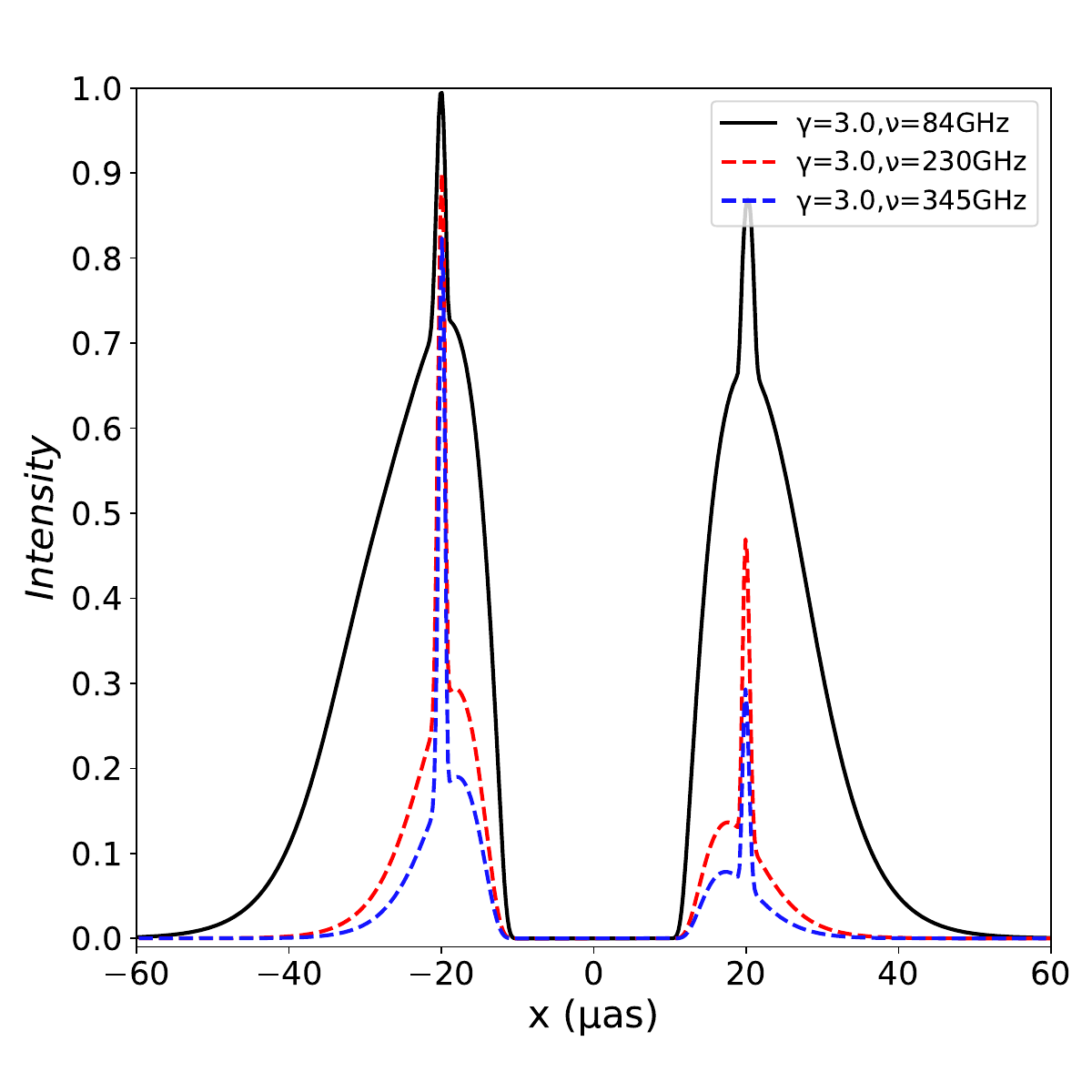}  
\includegraphics[width=2.32in]{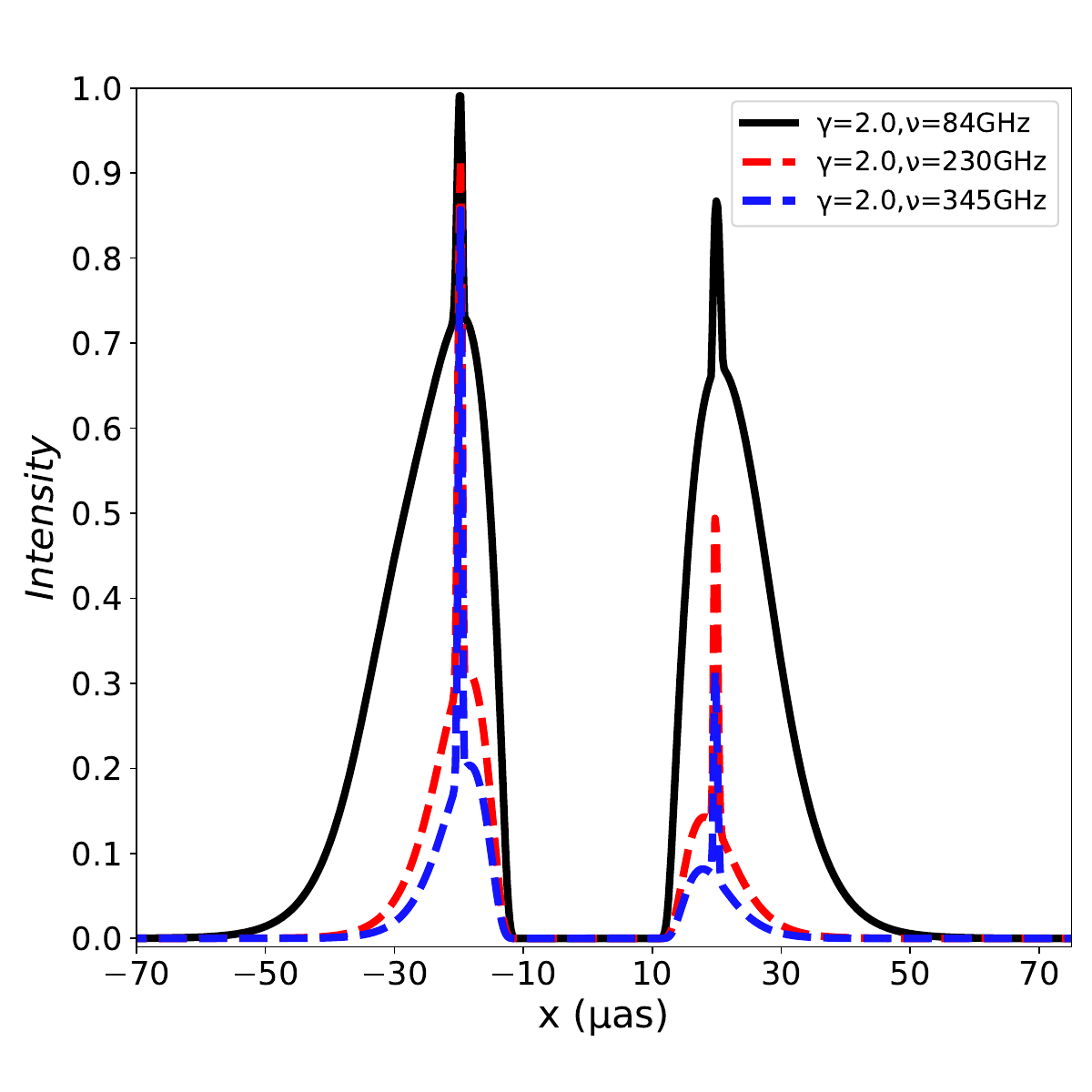}
\caption{The normalized intensity profiles along the horizontal cross section $y=0$ for the model images of Fig.\ref{imagesv} for different tidal parameters$\gamma$ and observed frequencies $\nu$.}
\label{imagesii}
\end{figure}

The Event Horizon Telescope (EHT) currently operates at 230 GHz, with ongoing efforts to expand its frequency coverage for both the current EHT and the next-generation EHT (ngEHT). Accordingly, we have simulated the RIAF images of braneworld black holes across a broad spectrum of frequencies for different tidal parameter $\gamma$. Fig. \ref{imagesv} clearly demonstrates that the emission intensity decreases  and the central dark region of the image slightly increases with the observational frequency, which is also shown in Fig. \ref{imagesii}. Furthermore, with the increase of $\gamma$, the primary and secondary peak values increases at the low observed frequency $\nu=84 GHz$ and decrease at the high observed frequency $\nu=230 GHz$ and $\nu=345GHz$, while the peak width is monotonically decreasing at these observed frequencies.
\begin{figure}
\centering
\includegraphics[width=7in]{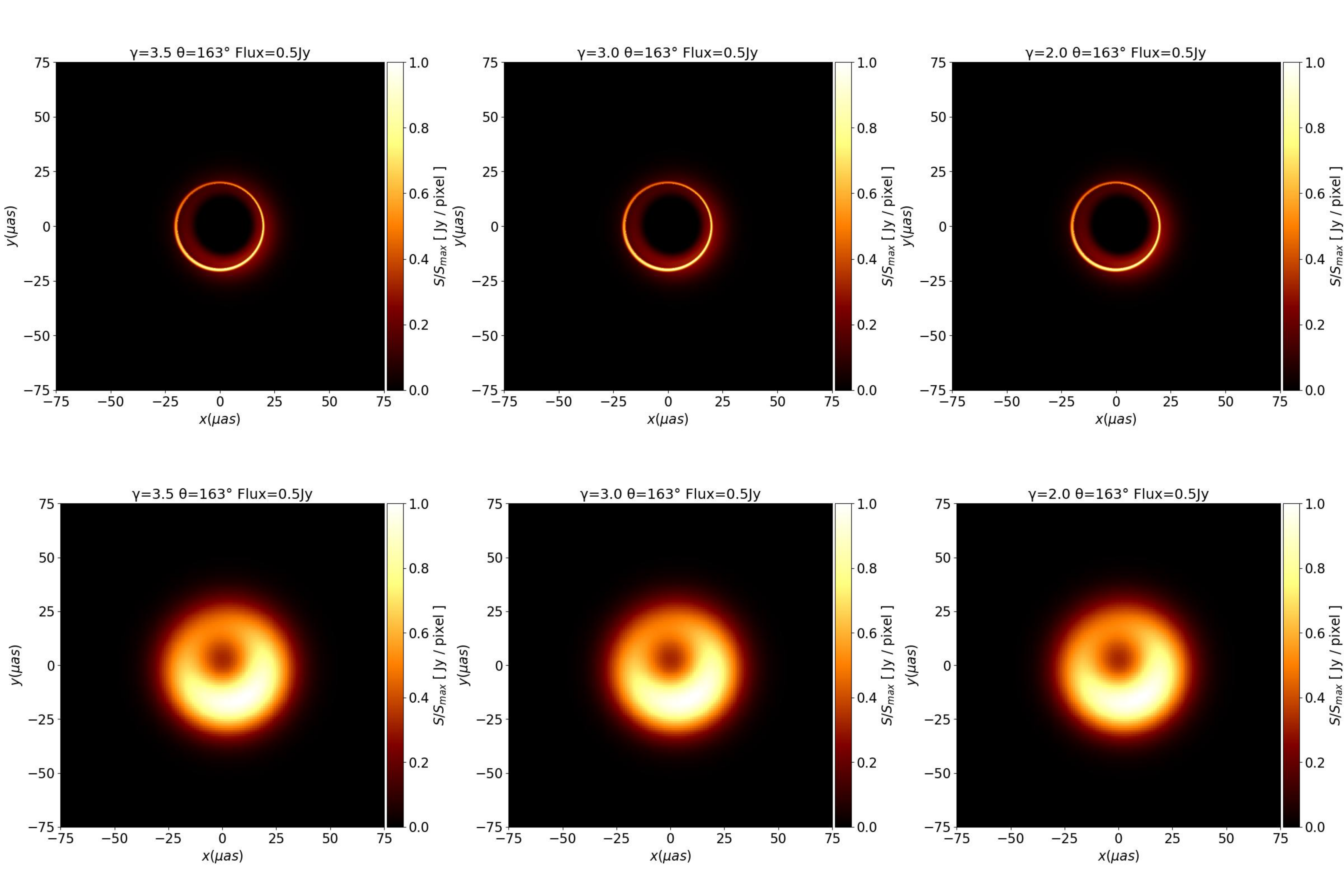} 
\caption{Black hole images (in the top row) and the corresponding blurred images with a half width  $20\mu as$ (in the bottom row)  at an inclination angle of $\theta=163^\circ$ and a position angle $PA=288^\circ$.}
\label{imagesg}
\end{figure}

To facilitate a comparison with the M87* observations, in Fig. \ref{imagesg} we present simulated images of the RIAF around braneworld black holes, together with the corresponding blurred images with a FWHM of $20\mu as$, at an inclination angle $\theta=163^{\circ}$  and position angle $PA=288^{\circ}$ \cite{Cui:2023uyb,CraigWalker:2018vam}. A Gaussian beam with $20\mu as$ FWHM is adopted to account for the finite angular resolution of the Event Horizon Telescope \cite{EventHorizonTelescope:2022wkp,EventHorizonTelescope:2021bee,Mann:2018xkm}. These simulated images exhibit morphological consistency with the current EHT observations of M87*.  However, the difference between the images of the braneworld black hole and the Schwarzschild black hole is too subtle to be resolved by current instrumentation  after applying $20\mu as$ Gaussian blurring.  It is natural to analyze whether these tiny differences can be detected by future experimental measurements with technological developments. The normalized cross-correlation coefficient (nCCC) can be used to  quantify the ``overlap" between two images  $ I $ and $K $ by comparing their pixel-wise differences.  The value $\text{nCCC} = 1$ refers to two identical images, while $\text{nCCC} = 0$ indicates no correlation between these images.\cite{Uniyal:2025uvc}
\begin{figure}
\centering
\includegraphics[width=3.4in]{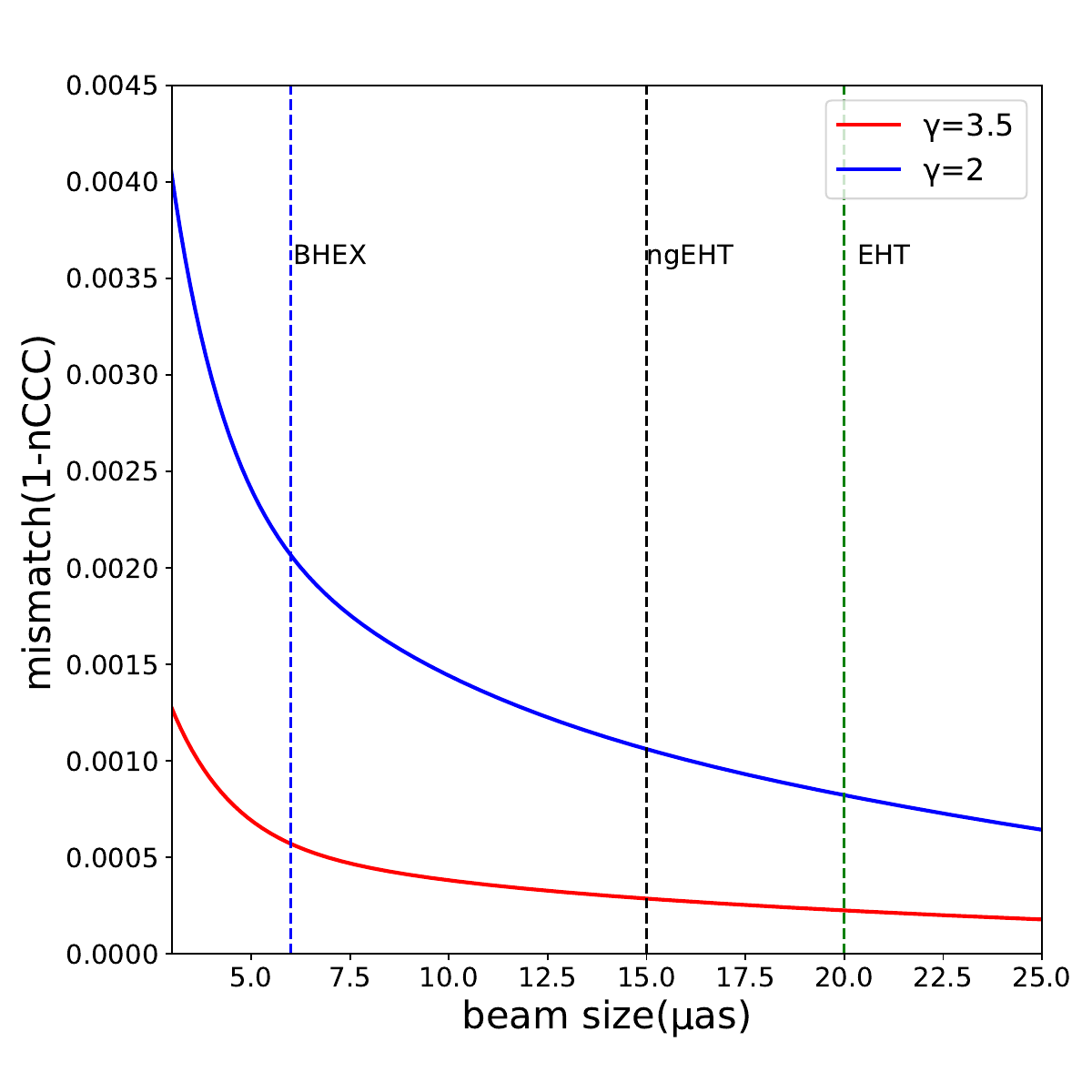} \includegraphics[width=3.6in]{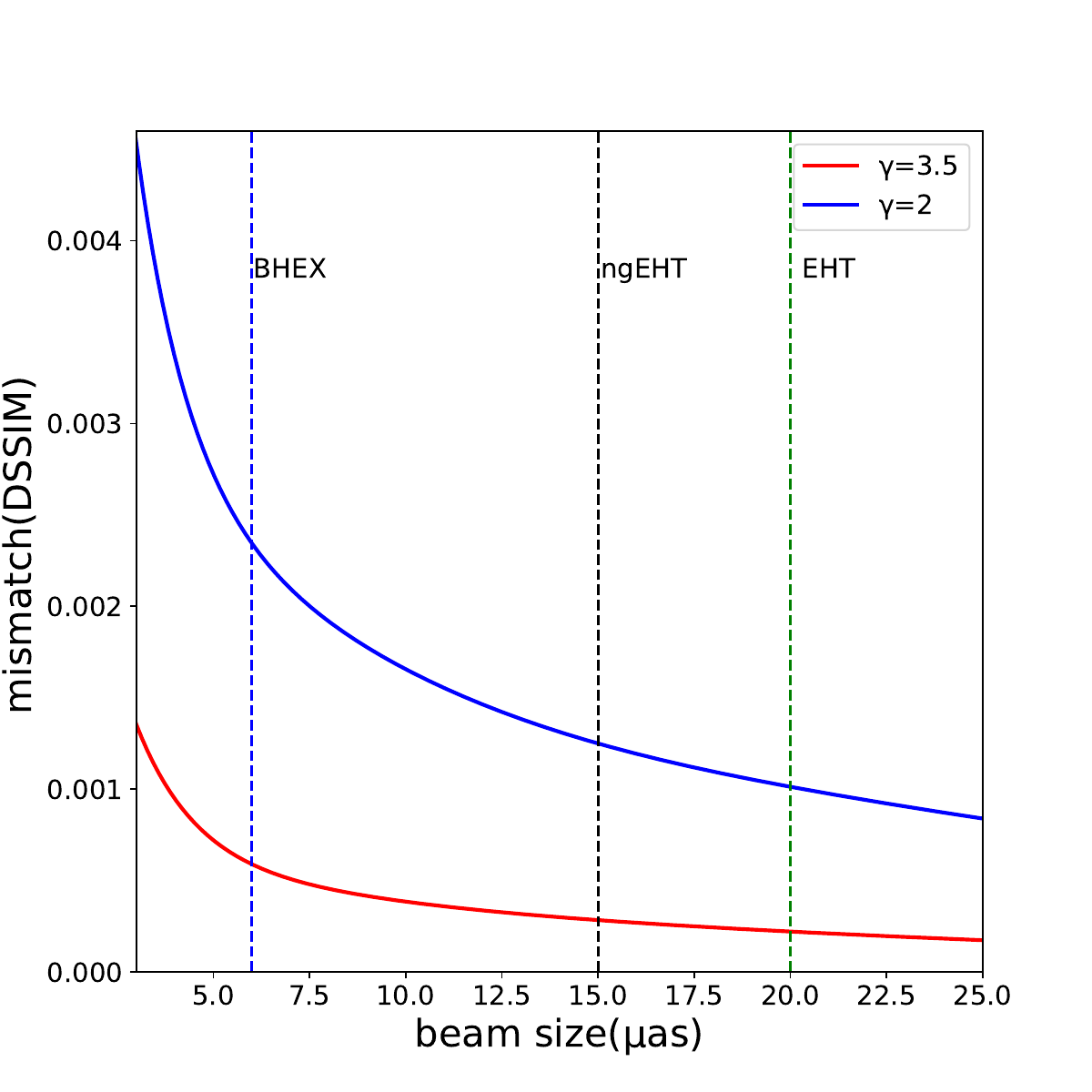} 
\caption{ Changes of image-comparison metrics between braneworld black hole and Schwarzschild black hole with beam size. The left panel is for the ``mismatch" $1-\rm{nCCC}$ and the right panel is for the DSSIM index. The vertical lines denote the present EHT resolutions and the expected ones for ngEHT and BHEX, respectively.}
\label{imagesC}
\end{figure}
Then, the ``mismatch" $ 1 - \text{nCCC}$ can be calculated by\cite{Wang:2004ssim}
\begin{eqnarray}
1 - \text{nCCC}(I, K) := 1 - \frac{1}{N} \sum_{i} \frac{(I_i - \mu_I)(K_i - \mu_K)}{\sigma_I \sigma_K},\label{mismatch}
\end{eqnarray}
where $ \mu_I $ and $ \mu_K $ are the mean pixel values in the two images $ I $ and $K $, and $ \sigma_I $ and $ \sigma_K $ are the standard deviations of the pixel values for the two images,
\begin{eqnarray}
\mu_I=  \sum_{i} \frac{I_i}{N},\quad\quad\quad \sigma^2_I=\frac{1}{N-1}\sum_{j=1}^{N} (I_j - \mu_j)^2.
\end{eqnarray}
The sum is taken over all $ N $ pixels in both images.

The structural dissimilarity (DSSIM) index serves as an alternative metric for quantifying the discrepancy between two images, which offers a different perspective on image variance. The DSSIM can be given by\cite{Mizuno:2018lxz}
\begin{eqnarray}
\text{DSSIM}(I, K) = \left|\frac{\sigma_I^2 + \sigma_K^2}{2 \sigma_{IK}} \right| -1,
\label{dssim}  
\end{eqnarray}
with
\begin{eqnarray}
\sigma_{IK}=\frac{1}{N-1}\sum_{j=1}^{N} (I_j - \mu_I)(K_j - \mu_K).
\end{eqnarray}
Fig.\ref{imagesC} shows the image-comparison metric in terms of the ``mismatch" $1-\rm {nCCC}$ and  the DSSIM index for different beam sizes between braneworld black hole and Schwarzschild black hole ($\gamma=3.0$)  with an internal resolution  $\sim0.25\mu as$, which is much smaller than that of the future projects considered. 
The blue and red curves represent the evolution of the mismatch between the Schwarzschild black hole and the braneworld black hole with the tidal parameter $\gamma=3.5$ and $\gamma=2$, respectively. It is shown that both the mismatche and  the DSSIM index decrease monotonically with the beam size. It is reasonable because a larger beam size implies a lower angular resolution. The vertical dashed lines in green, black, and blue denote the angular resolution of the current EHT, as well as the projected resolutions of the future ngEHT and BHEX, respectively. 
It is clear that increasingly precise experiments will enable a more robust distinction between braneworld black hole and the Schwarzschild black hole.
Moreover, we find that the discrepancy in images between the $\gamma=2$ braneworld black hole and the Schwarzschild black hole is slightly larger than that for the $\gamma=3.5$ case.  For the braneworld black holes, we also note that the mismatch from the DSSIM index is overall larger than that reported from the nCCC metric, i.e., $\rm{DSSM/(1-nCCC)}\sim 1.1$, which is different from that of Konoplya-Rezzolla-Zhidenko (KRZ) black holes $\rm{DSSM/(1-nCCC)}\sim 1/3$. However, Fig. \ref{imagesC} implies that the magnitudes of these mismatches are on the order of $10^{-3}$, which means that the identification of braneworld black holes through black hole images is more challenging than that of KRZ black holes.

\section{Discussions and conclusions}
\label{Conclusion}

We have investigated images illuminated by thermal synchrotron emission from radiatively inefficient accretion flows around of braneworld black holes. Our results reveal that  effects of the tidal parameter $\gamma$ on black hole images depends on the inclination angle $\theta$, the disk thickness $h$ and the observed frequency. The primary peak value of the luminosity distribution decreases with the parameter $\gamma$ for $\theta=45^{\circ}$ and $\theta=163^{\circ}$, but  increases slightly for $\theta=90^{\circ}$. The secondary peak value  decreases with the parameter $\gamma$ for $\theta=163^{\circ}$, whereas for $\theta=45^{\circ}$ and $\theta=90^{\circ}$, it first decreases and then increases. The peak width generally increases with the parameter $\gamma$, except for the secondary peak at $\theta=45^{\circ}$, which exhibits a decreasing trend. For a fixed inclination $\theta$, the inner shadow diameter decreases with the parameter $\gamma$.  At a fixed inclination $\theta=163^{\circ}$, the primary and secondary peak values and their width decrease with the increase of $\gamma$ for different disk thickness $h$. With the increase of $h$, both the intension peak value  and the inner shadow radius decrease. Furthermore, as the parameter $\gamma$ increases, the primary and secondary peak values increases at the low observed frequency $\nu=84 GHz$ and decrease at the high observed frequency $\nu=230 GHz$ and $\nu=345GHz$, while the peak width is monotonically decreasing at these observed frequencies.

Finally, we analyze the image-comparison metric in terms of the ``mismatch" $1-\rm {nCCC}$ and  the DSSIM index for different beam sizes between braneworld black hole and Schwarzschild black hole. It is found that the discrepancy in images between the $\gamma=2$ braneworld black hole and the Schwarzschild black hole is slightly larger than that for the $\gamma=3.5$ case.  We also note that for the braneworld black holes, the mismatch from the DSSIM index is overall larger than that reported from the nCCC metric, i.e., $\rm{DSSM/(1-nCCC)}\sim 1.1$, which is different from the corresponding ratio for KRZ black holes $\rm{DSSM/(1-nCCC)}\sim 1/3$. However, given that the magnitudes of these mismatches are on the order of $10^{-3}$,  the identification of braneworld black holes through black hole images is more challenging than that of KRZ black holes.

\section*{Acknowledgements}
This work was supported by the National Natural Science Foundation of China (Grant Nos. 12275079 and 12035005), the National Key Research and Development Program of China (Grant No. 2020YFC2201400), and the innovative research group of Hunan Province under Grant No. 2024JJ1006.

\bibliography{references}

\end{document}